\documentclass[11pt,journal,onecolumn]{IEEEtran}
\usepackage{amsmath,amssymb,bm,amsthm,amstext}
\usepackage{enumerate,longtable,tikz,subfigure,float}
\usepackage{cases}
\usepackage{bbm}
\usepackage{cite,color}
\usepackage{multirow}
\usetikzlibrary{decorations.pathreplacing}
\usepackage{cleveref}

\crefname{lem}{Lemma}{Lemma}
\crefname{thm}{Theorem}{Theorem}
\crefname{dfn}{Definition}{Definition}
\crefname{prop}{Proposition}{Proposition}
\crefname{example}{Example}{Example}
\crefname{rmk}{Remark}{Remark}
\crefname{cor}{Corollary}{Corollary}
\crefname{clm}{Claim}{Claim}

\newtheorem{thm}{Theorem}[section]
\newtheorem{lem}{Lemma}[section]
\newtheorem{dfn}{Definition}[section]
\newtheorem{prop}{Proposition}[section]
\newtheorem{rmk}{Remark}[section]
\newtheorem{cor}{Corollary}[section]
\newtheorem{example}{Example}[section]
\newtheorem{clm}{Claim}

\newenvironment{pf}{{\it\textbf{Proof of Lemma VI.1}}}{\hfill $\blacksquare$\par}
\newcommand{\Rnum}[1]{\lowercase\expandafter{\romannumeral #1\relax}}
\newcommand{\RNum}[1]{\uppercase\expandafter{\romannumeral #1\relax}}
\newcommand{\Inv}[1]{\textup{Inv}\left(#1\right)}
\newcommand{\tabincell}[2]{\begin{tabular}{@{}#1@{}}#2\end{tabular}}

\title{ Reconstruction of Sequences Distorted by Two Insertions}
\author{Zuo~Ye, Xin~Liu, Xiande~Zhang and Gennian~Ge%
\thanks{This project was supported by the National Key Research and Development Program of China under Grant 2020YFA0712100,  Grant 2018YFA0704703 and Grant 2020YFA0713100, the National Natural Science Foundation of China under Grant 11971325, Grant 12171452 and Grant 12231014,  and Beijing Scholars Program.}

\thanks{Z. Ye and X. Zhang are with  the School of Mathematical Sciences,
	University of Science and Technology of China, Hefei 230026, Anhui, China
	(e-mails: zyprince@mail.ustc.edu.cn; drzhangx@ustc.edu.cn).}%
\thanks{X. Liu and G. Ge are
	with  the School of Mathematical Sciences, Capital Normal University,
	Beijing 100048, China (e-mails: liuxin\_2020@yeah.net;
	gnge@zju.edu.cn).}%
}

\begin{document}
\maketitle

\begin{abstract}
Reconstruction codes are generalizations of error-correcting codes that can correct errors by a given number of noisy reads. The study of such codes was initiated by Levenshtein in 2001 and developed recently due to applications in modern storage devices such as racetrack memories and DNA storage. The central problem on this topic is to design codes with redundancy as small as possible for a given number $N$ of noisy reads. In this paper, the minimum redundancy of such codes for binary channels with exactly two insertions is determined asymptotically for all values of $N\ge 5$.
Previously, such codes were studied only for channels with single edit errors or two-deletion errors.
\end{abstract}

\section{Introduction}
The study of sequence reconstruction problems was initiated by Levenshtein in \cite{wz6,wz7,wz8} to combat
errors by repeatedly transmitting a message without coding, which is of interest in fields like informatics, molecular biology, and chemistry.
The setting of this problem is as follows: the sender transmits a sequence 
through $N$ different noisy channels and the receiver receives all the  channel outputs, called noisy \emph{reads}; then the receiver aims to reconstruct the transmitted sequence with the help of these $N$ outputs. There are probabilistic and combinatorial versions of this problem. In the probabilistic version, the noisy reads are obtained by  transmitting the sequence $\bm{x}$ through probabilistic channels, and the goal is to minimize the value of $N$ required for reconstructing $\bm{x}$ with high probability, see \cite{wz7,proc2,proc3,proc4,proc5,wz15,proc7} and the references therein. In the combinatorial version, noisy reads are obtained by introducing the maximum number of errors, and the goal is to determine the minimum value of $N$ needed for zero-error reconstruction \cite{wz7}.

In this paper, we focus on the combinatorial version,
which requires that all channel outputs are different from each other\cite{wz7}. Then the  problem of determining the minimum value of $N$ is reduced to determining the maximum intersection size of two distinct error balls. Specifically, the maximum intersection size is equal to $N-1$.  In \cite{wz7,wz8}, Levenshtein solved the problem for several error types, such as substitutions, insertions, deletions, transpositions, and asymmetric errors. In \cite{wz16,wz17,wz18}, permutation errors were analyzed. For general error graphs, see \cite{wz7,wz19,wz20}. Note that these works considered the uncoded case, that is, the transmitted sequences are selected from the entire space. Due to applications in DNA storage, many researchers also investigated this problem under the setting where the transmitted sequences are chosen from a given code with a certain error-correcting property, see for example \cite{proc8,proc9,wz11,wz1,wz27}.

Recently, the dual problem of the sequence reconstruction was developed under the scenario that the number of noisy reads is a fixed system parameter.
Motivated by applications in DNA-based data storage and racetrack memories, Kiah et al. \cite{proc6,wz2} and Chrisnata et al. \cite{wz13} initiated the study of designing \emph{reconstruction codes}, for which  the  size of intersections between any two error-balls is strictly less than the number of reads $N$, for a given $N$. Note that when $N=1$, reconstruction codes are the classical error-correcting codes. When $N>1$,
one can increase the information capacity, or equivalently, reduce the number of redundant
bits by leveraging on these multiple  reads. The redundancy of a binary code $\mathcal{C}$ of length $n$ is defined to be $n - \log_2 | \mathcal{C} |$. In \cite{proc6,wz2}, the authors  focused on channels that introduce single edit error, that is a single substitution, insertion, or deletion, and their variants. For  single deletion binary codes, they showed that the number of redundant symbols required can be reduced from
$\log_2(n)+O(1)$ to $\log_2\log_2(n)+O(1)$ if one increases the number of noisy reads $N$ from one to two. 
For two deletions, the best known $2$-deletion correcting code (i.e., $N=1$) has redundancy $4\log_2(n)+o(\log_2(n))$ \cite{wz4}. In \cite{wz13,wz28}, Chrisnata et al. reduced this redundancy to $2\log_2(n)+o(\log_2(n))$ by increasing the number
of noisy reads $N$ from one to five, while each noisy read  suffers exactly two deletions. For $t$-deletions (with $t\geq 2$), it was proved in \cite{wz28} that the reconstruction code with two reads for a single deletion \cite{wz2} is able to reconstruct codewords from $N=n^{t-1}/(t-1)!+O(n^{t-2})$
distinct noisy reads.

It is well known that a code can correct $t$-deletions (i.e., $N=1$) if and only if it can correct $t$-insertions. However, it is not the case when $N>1$. When $N>1$, the problem is closely related to the size of the intersection of two error balls. From \cite{wz11}, \cite{wz1} and \cite[Proposition 9]{wz2}, we can see that in general the intersection size of two $t$-deletion balls is different from that of two $t$-insertion balls.

In this paper, we study binary reconstruction codes for channels that cause exactly two insertion errors.
We show that the number of redundant symbols required to reconstruct a codeword uniquely can be reduced from $4\log_2(n)+o(\log_2(n))$ to $\log_2(n)+O(\log_2\log_2(n))$ by increasing the number of noisy reads $N$ from one to five. The redundancy can be further reduced to $\log_2\log_2(n)+\Theta(1)$ if the number of noisy reads is at least $n+4$. For readers' easy reference, we summarize the best known results on binary $2$-deletion reconstruction codes and $2$-insertion reconstruction codes in \Cref{tb_comparison}. For general  $t$-insertions (with $t> 2$), we can show the similar result as in \cite{wz28}: the reconstruction code with two reads for a single insertion \cite{wz2} is able to reconstruct codewords from $N=n^{t-1}/(t-1)!+O(n^{t-2})$ distinct noisy reads.

\begin{table}
  \centering
  \caption{Best known results, where \textbf{DRC} denotes deletion reconstruction codes and \textbf{IRC} denotes insertion reconstruction codes}\label{tb_comparison}
  \begin{tabular}{c|c|c|c}
    \hline
    &The value of $N$&Redundancy&Reference\\
    \hline
    \multirow{5}{*}{$2$-\textbf{DRC}}&$N\le 4$&$\Theta(\log_2(n))$&\cite[Theorem 1]{wz4}\\
    \cline{2-4}
    &$N=5$&$2\log_2(n)+O(\log_2\log_2(n))$&\cite[Theorem 10]{wz28}\\
    \cline{2-4}
    &$N=7$&$\log_2(n)+O(1)$&\cite[Theorem 5]{wz28}\\
    \cline{2-4}
    &$N=n+1$&$\log_2\log_2(n)+O(1)$&\cite[Theorem 9]{wz28}\\
    \cline{2-4}
    &$N>2n-4$&$0$&\cite[Page 7]{wz7},\cite[Theorem 3]{wz28}\\
    \hline
    \multirow{5}{*}{$2$-\textbf{IRC}}&$N\le 4$&$\Theta(\log_2(n))$&\cite[Theorem 1]{wz4}\\
    \cline{2-4}
    &$N=5,6$&$\log_2(n)+O(\log_2\log_2(n))$&\Cref{thm_construction4}\\
    \cline{2-4}
    &$6<N\le n+3$&$\log_2(n)+\Theta(1)$&\Cref{prop_summary1}\\
    \cline{2-4}
    &$n+4\le N\le 2n+4$&$\log_2\log_2(n)+\Theta(1)$&\Cref{prop_summary1},\Cref{thm_construction1}\\
    \cline{2-4}
    &$N>2n+4$&$0$&\Cref{prop_summary1}\\
    \hline
  \end{tabular}
\end{table}

This paper is organized as follows. In \Cref{sec_preliminary}, we introduce some notations and the problem statement formally. In \Cref{sec_first}, we derive some preliminary results, which help us to solve the trivial cases. In \Cref{sec_np4np5}, we first completely characterize the structures of two sequences when the intersection size of their $2$-insertion balls is exactly $n+4$ or $n+5$. Then by utilizing the structures, we give explicit reconstruction codes for $N=n+4$, which are asymptotically optimal in terms of redundancy. In \Cref{sec_N5}, by applying higher order parity checks on short subwords, we construct asymptotically optimal codes when $N=5$. Finally, we conclude this paper in \Cref{sec_conclusion} with some open problems.

\section{Notations and Problem statement}\label{sec_preliminary}
In this section, we introduce some notations and preliminary results needed throughout this paper.
Let $\Sigma_2$ denote the alphabet $\{0,1\}$, and $\Sigma_2^n$ denote the set of all binary sequences of length $n$. Further, let $\Sigma_2^{*}\triangleq\mathop{\cup}\limits_{n=0}^{\infty}\Sigma_2^n$, that is the set consisting of all binary sequences of finite length. Here if $n=0$, we mean the empty sequence, denoted by $\emptyset$.

For a sequence $\bm{x}=x_1\cdots x_n\in\Sigma_2^n$ and a sequence $\bm{z}=z_1\cdots z_t\in\Sigma_2^t$, where $t$ is an integer satisfying $1\le t\le n$, if there exist $1\le i_1<\cdots <i_t\le n$ such that $z_j=x_{i_j}$ for all $1\le j\le t$, we say that $\bm{z}$ is a \textit{subsequence} of $\bm{x}$, or $\bm{x}$ is a \textit{supersequence} of $\bm{z}$.
In particular, when $i_{j+1}=i_j+1$ for all $j$, we say  that $\bm{z}$ is a \textit{subword} of $\bm{x}$, and use the notation $\bm{x}[l:k]$ with $l=i_1$ and $k=i_t$ to denote $\bm{z}$. Let $|\bm{x}|$ denote the length of $\bm{x}$.
For any $\bm{x}\in\Sigma_2^n$ and $t\ge 0$, let $I_t(\bm{x})\triangleq\{\bm{z}\in\Sigma_2^{n+t}:\text{ }\bm{z} \text{ is a supersequence of } \bm{x}\}$. We call $I_t(\bm{x})$ the \emph{$t$-insertion ball} centered at $\bm{x}$. It is known that $\left|I_t(\bm{x})\right|$ is independent of the choice of $\bm{x}$ by \cite{wz7} and equal to
\begin{equation}\label{Nl=0}
  I(n,t)\triangleq \mathop{\sum}\limits_{i=0}^{t}\binom{n+t}{i}.
\end{equation}
 The \emph{Levenshtein distance} between $\bm{x}$ and $\bm{y}$ in $\Sigma_2^n$, denoted by $d_L(\bm{x},\bm{y})$, is the smallest integer $\ell$ such that $I_{\ell}(\bm{x})\cap I_{\ell}(\bm{y})$ is not empty. For example, $d_L(10100,01001)=1$.
It is easy to see that $0\le d_L(\bm{x},\bm{y})\le n$, and $d_L(\bm{x},\bm{y})=0$ if and only if $\bm{x}=\bm{y}$. Finally,  the Hamming distance between $\bm{x}$ and $\bm{y}$, denoted by $d_H(\bm{x},\bm{y})$, is the number of coordinates where $\bm{x}$ and $\bm{y}$ differ, and the Hamming weight $wt_H(\bm{x})$ of $\bm{x}$ is the number of nonzero coordinates of $\bm{x}$.

The intersection size of two error balls is a key ingredient of the reconstruction problem in coding theory. By Levenshtein's original framework \cite{wz7}, for channels causing $t$ insertion errors, it is always possible to exactly reconstruct $\bm{x}\in\Sigma_2^n$ given $N^{+}(n,t)+1$ distinct elements of $I_t(\bm{x})$, where 
\begin{equation}\label{Nl=1}
\begin{aligned}
   N^{+}(n,t)&\triangleq\max\{\left| I_t(\bm{x})\cap I_t(\bm{y}) \right|:\text{ }\bm{x},\bm{y}\in\Sigma_2^n\text{ and }\bm{x}\neq \bm{y}\}\\
  &=\mathop{\sum}\limits_{i=0}^{t-1}\binom{n+t}{i}(1-(-1)^{t-i}).
\end{aligned}
\end{equation}
This notation was  generalized
in \cite{wz1} for any two sequences with a minimum Levenshtein distance. For integers $n$, $t\ge \ell\ge 0$, let
$$
N^{+}(n,t,\ell)\triangleq\max\{\left| I_t(\bm{x})\cap I_t(\bm{y}) \right|:\text{ }\bm{x},\bm{y}\in\Sigma_2^n\text{ and }d_L(\bm{x},\bm{y})\ge \ell\},
$$ then the explicit formula is given by

\begin{equation}\label{comN}
 N^{+}(n,t,\ell)=\mathop{\sum}\limits_{j=\ell}^{t}\mathop{\sum}\limits_{i=0}^{t-j}\binom{2j}{j}\binom{t+j-i}{2j}\binom{n+t}{i}(-1)^{t+j-i}.
\end{equation}
Note that when $\ell=0,1$, the values of $N^{+}(n,t,l)$ reduce to $I(n,t)$ and $N^{+}(n,t)$, respectively. See \cite[Corollary 9]{wz1} and \cite[Corollary 10]{wz1} for details.

Given a code $\mathcal{C}\subseteq\Sigma_2^n$ with $|\mathcal{C}|\ge2$,  the \textit{read coverage} $\nu_t (\mathcal{C})$ of $\mathcal{C}$ after $t$ insertions is defined as
\begin{equation}\label{eq_readcoverage}
  \nu_t (\mathcal{C})\triangleq\max\{\left| I_t(\bm{x})\cap I_t(\bm{y}) \right|:\text{ }\bm{x},\bm{y}\in\mathcal{C}\text{ and }\bm{x}\ne \bm{y}\}.
\end{equation}
 It is clear that if $\nu_t (\mathcal{C})<N$, we can uniquely recover any codeword in $\mathcal{C}$ by its $N$ distinct reads. So we have the following definition.
\begin{dfn}\textup{(Reconstruction codes)}\label{dfn_reconstructioncode}
A code $\mathcal{C}\subset \Sigma_2^n$ is an $(n,N;B^{I(t)})$-\emph{reconstruction code} if  $\nu_t (\mathcal{C})<N$, where $\nu_t (\mathcal{C})$ is the read coverage of $\mathcal{C}$ after $t$ insertions.
\end{dfn}
Given a  code $\mathcal{C}\subset \Sigma_2^n$, the \emph{redundancy} of $\mathcal{C}$ is defined to be the value $n-\log_2\left|\mathcal{C}\right|$. Then we are interested in the  following quantity
$$
\rho(n,N;B^{I(t)})\triangleq\min\{n-\log_2\left|\mathcal{C}\right|:\text{ }\mathcal{C}\subseteq\Sigma_2^n \text{ and } \nu_t(\mathcal{C})<N\},
$$
which is called the \emph{optimal} redundancy of an $(n,N;B_2^{I(t)})$-reconstruction code. 
By definition, an $(n,N;B^{I(t)})$-reconstruction code is also an $(n,N^{\prime};B^{I(t)})$-reconstruction code for any  $N^{\prime}>N$. So we have $\rho(n,N;B^{I(t)})\geq \rho(n,N+1;B^{I(t)})$ for all $N\geq 1$.

When $N=1$, an $(n,1;B^{I(t)})$-reconstruction code is indeed a $t$-insertion correcting code, or equivalently a $t$-deletion correcting code. 
This class of codes has been extensively studied in recent years, see \cite{wz22,wz23,wz5,wz24,wz4} for more details. For $t=1$, we have the following single deletion correcting code \cite{wz9}:
\begin{equation}\label{eq_VTcode}
\left\{\bm{x}=x_1\cdots x_n\in\Sigma_2^{n}\;:\;\mathop{\sum}\limits_{i=1}^{n}ix_i\equiv a\pmod{n+1}\right\},
\end{equation}
where $a$ is a fixed integer between $0$ and $n$. This is the famous Varshamov-Tenengolts (VT) code, which has asymptotically optimal redundancy $\log_2(n+1)$ when $a=0$ \cite[Corollary 2.3]{Sloane2002}. In \cite[Theorem 24]{wz2}, Cai et al.  determined the asymptotic value of $\rho(n,N;B^{I(1)})$ for all $N\ge 1$ as below.


\begin{thm}\label{thm_1inser}
  Taking $q=2$ in \cite[Theorem 24]{wz2}, we have
  \begin{equation*}
    \rho(n,N;B^{I(1)})=
    \begin{cases}
      \log_2(n)+\Theta(1), & \mbox{if } N=1, \\
      \log_2\log_2(n)+\Theta(1), & \mbox{if } N=2, \\
      0, & \mbox{if }N\ge 3.
    \end{cases}
  \end{equation*}
\end{thm}

Our work  is to extend the result in Theorem~\ref{thm_1inser} to the case of $t=2$. As will be shown in \Cref{cor_n5} and the paragraph prior to \Cref{prop_summary1}, the lower bound of the optimal redundancy can be deduced from the case $t=1$. The upper bound needs explicit code constructions. So the main task is to construct an $(n,N;B^{I(2)})$-reconstruction code with  redundancy as small as possible, for any given $N$.
Before closing this section, we introduce the following notations which will be used later.

For $\bm{x},\bm{y}\in\Sigma_2^{*}$, we write $\bm{x}\bm{y}$ for the concatenation of $\bm{x}$ and $\bm{y}$. In particular, if $i\ge 1$, we define $\bm{x}^{i}$ to be the sequence obtained by concatenating $i$ $\bm{x}$'s; if $i=0$,  $\bm{x}^{0}$ is the empty sequence. If $\bm{u}=u_1\cdots u_n\in\Sigma_2^{n}$, we denote $\overline{\bm{u}}=(1-u_1)\cdots (1-u_n)$ as the complementary sequence of $\bm{u}$.  For any $S\subseteq\Sigma_2^{*}$ and $a,b\in\Sigma_2$, we denote $S^a\triangleq\{\bm{x}\in S:\text{ }\bm{x} \text{ starts with }a\}$, $S_{b}\triangleq\{\bm{x}\in S:\text{ }\bm{x} \text{ ends with }b\}$, $S_{b}^a\triangleq\{\bm{x}\in S:\text{ }\bm{x}\text{ starts with }a \text{ and ends with }b\}$, $aS\triangleq\{a\bm{y}:\text{ }\bm{y}\in S\}$, and $Sb\triangleq\{\bm{y}b:\text{ }\bm{y}\in S\}$.
\begin{example}
 Let $\bm{x}=10$, $\bm{y}=101$ and $\bm{u}=101101$. Then $\bm{x}\bm{y}=10101$, $\bm{x}^2=1010$ and $\overline{\bm{u}}=010010$. Let $S=\left\{10101,0101,01010\right\}$, $a=0$ and $b=1$. Then $S^a=\left\{0101,01010\right\}$, $S_b=\left\{10101,0101\right\}$, $S_b^a=\left\{0101\right\}$, $aS=\left\{010101,00101,001010\right\}$ and $Sb=\left\{101011,01011,010101\right\}$.
\end{example}

Given a sequence $\bm{x}=x_1\cdots x_n\in\Sigma_2^{n}$ with $n\ge 2$, we say that  $\bm{x}$ has
 period $\ell$ ($\leq n$) if $\ell$ is the smallest integer such that  $x_i=x_{i+\ell}$ for all $1\leq i\leq n-\ell$. A sequence with period $2$ is called  \emph{alternating}. We also view an empty sequence or a length-one sequence as an alternating sequence.
 For example, $\emptyset$, $1$, $0$, $10$, $01$, $101$ are all alternating sequences.
%

\section{Preliminary results}\label{sec_first}

As mentioned before, the intersection size of two error balls is a key ingredient of the reconstruction problem. In this section, we characterize the structures of two binary sequences whose intersection of $2$-insertion balls are of a certain size.

Let us first recall the results for a single insertion, that is the size of  $I_{1}(\bm{x})\cap I_{1}(\bm{y})$. By Equation~(\ref{Nl=1}),
we know that $\left|I_{1}(\bm{x})\cap I_{1}(\bm{y})\right|\le 2$ for any two distinct sequences $\bm{x},\bm{y}\in\Sigma_2^n$. The characterization of $\bm{x},\bm{y}$ when the equality holds was given in \cite{wz2}, which needs the following notations.


\begin{dfn}\label{df_typA}\textup{(Type-A confusability)}
  Suppose that $\bm{x}\neq\bm{y}\in\Sigma_2^n$. We say that $\bm{x},\bm{y}$ are \emph{Type-A confusable}, if there exist $\bm{u}$, $\bm{v}$, $\bm{w}$ $\in\Sigma_2^{*}$ such that
 $$\bm{x}=\bm{u}\bm{w}\bm{v} \text{ and }\bm{y}=\bm{u}\overline{\bm{w}}\bm{v},$$
where $\bm{w}$ is an alternating sequence of length at least one, i.e., $\bm{w}=(a\bar{a})^i$ for some $i\ge 1$ or $\bm{w}=(a\bar{a})^ja$ for some $j\ge 0$, where $a=0$ or $1$.
\end{dfn}

For example, the two sequences $110$ and $100$ are Type-A confusable with $\bm{u}=1$, $\bm{w}=1$ and $\bm{v}=0$. In general, if the Hamming distance $d_H(\bm{x},\bm{y})=1$, then $\bm{x}$ and $\bm{y}$ are always Type-A confusable by definition. Note that the case where $d_H(\bm{x},\bm{y})=1$ was not included in the original definition in  \cite[Definition 8]{wz2}, where $j\geq 1$ is required when  $\bm{w}=(a\bar{a})^ja$. However, the following statement is still true,
which is just a straightforward combination of the two claims in \cite[Proposition 9]{wz2}.


\begin{lem}\label{lem_typA}\textup{\cite[Proposition 9]{wz2}}
  Let $\bm{x},\bm{y}\in\Sigma_2^n$ be distinct. Then $\left|I_{1}(\bm{x})\cap I_{1}(\bm{y})\right|$ $=2$ if and only if $\bm{x},\bm{y}$ are Type-A confusable.
\end{lem}

In  fact, if two sequences $\bm{x}$ and $\bm{y}$ are Type-A confusable, then $$
\left\{
\begin{array}{l}
\bm{x}=\bm{u}(a\bar{a})^{m+1}\bm{w}\\
\bm{y}=\bm{u}(\bar{a}a)^{m+1}\bm{w}
\end{array}
\right.
\text{\ or\ }
\left\{
\begin{array}{l}
\bm{x}=\bm{u}(a\bar{a})^{m}a\bm{w}\\
\bm{y}=\bm{u}(\bar{a}a)^{m}\bar{a}\bm{w}
\end{array}
\right.
$$ for some $\bm{u},\bm{w}\in\Sigma_2^{*}$, $a\in\Sigma_2$ and $m\ge 0$. By \Cref{lem_typA}, $I_1(\bm{x})\cap I_1(\bm{y})=\{\bm{u}(\bar{a}a)^{m+1}\bar{a}\bm{w},\bm{u}(a\bar{a})^{m+1}a\bm{w}\}$ for the left case and $I_1(\bm{x})\cap I_1(\bm{y})=\{\bm{u}(\bar{a}a)^{m+1}\bm{w},\bm{u}(a\bar{a})^{m+1}\bm{w}\}$ for the right case.
Next, we define the concept of Type-B confusability, which was introduced in \cite[Definition 18]{wz13} and \cite[Definition 11]{wz28}. This concept will help to characterize the case $\left|I_{1}(\bm{x})\cap I_{1}(\bm{y})\right|=1$.

\begin{dfn}\label{df_typB}\textup{(Type-B confusability)}
  Suppose that $\bm{x}\neq\bm{y}\in\Sigma_2^n$ with $n\geq 3$. We say that $\bm{x},\bm{y}$ are \emph{Type-B confusable}, if there exist $\bm{u}$, $\bm{v}$, $\bm{w}$ $\in\Sigma_2^{*}$ and $a,b\in\Sigma_2$ such that
  $$
  \left\{\bm{x},\bm{y}\right\}=\left\{\bm{u}a\bar{a}\bm{v}b\bm{w},\bm{u}\bar{a}\bm{v}b\bar{b}\bm{w}\right\}.
  $$
\end{dfn}
\begin{example}\label{examp_typAtypB}
We give some examples of these two types of confusability.
\begin{itemize}
  \item Let $\bm{x}=11101010$ and $\bm{y}=11010110$. Then $\bm{x}$ and $\bm{y}$ are Type-A confusable with $\bm{u}=11$, $\bm{w}=1010$ and $\bm{v}=10$. They are also Type-B confusable with $\bm{u}=11$, $\bm{v}=1$, $\bm{w}=10$, $a=1$ and $b=0$.
  \item Let $\bm{x}=111010$ and $\bm{y}=110110$. Then $\bm{x}$ and $\bm{y}$ are Type-A confusable with $\bm{u}=11$ and $\bm{w}=\bm{v}=10$. Obviously, they are not Type-B confusable.
  \item Let $\bm{x}=1110100110$ and $\bm{y}=1101001010$. Then $\bm{x}$ and $\bm{y}$ are Type-B confusable with $\bm{u}=11$, $\bm{v}=100$, $\bm{w}=10$ and $a=b=1$. It is easy to see that they are not Type-A confusable.
  \item The two sequences $0011$ and $1110$ are neither Type-A confusable nor Type-B confusable.
\end{itemize}
\end{example}

From \Cref{examp_typAtypB}, we can see that Type-A confusability and Type-B confusability do not contain each other. In general, we have the following proposition, which is easy to verify.
\begin{prop}\label{prop_summary}
From \Cref{df_typA} and \Cref{df_typB}, one can easily check that
\begin{itemize}
  \item[(1)] if $\bm{x}=\bm{u}\bm{w}\bm{v}$, $\bm{y}=\bm{u}\overline{\bm{w}}\bm{v}$ $\in\Sigma_2^n$ are Type-A confusable, then they are Type-B confusable if and only if
   $\bm{w}\in\{(10)^m,(01)^m\}$ for some $m\ge 2$, or $\bm{w}\in\{(10)^{m}1,(01)^{m}0\}$ for some $m\ge 1$;
  \item[(2)] if $\bm{x}=\bm{u}a\bar{a}\bm{v}b\bm{w}$, $\bm{y}=\bm{u}\bar{a}\bm{v}b\bar{b}\bm{w}$ $\in\Sigma_2^n$ are Type-B confusable, then they are Type-A confusable if and only if one of the following two conditions is satisfied:
      \begin{itemize}
        \item[(i)] $a=b$ and $\bm{v}=(a\bar{a})^m$ for some $m\ge 0$;
        \item[(ii)] $a=\bar{b}$ and $\bm{v}=(a\bar{a})^ma$ for some $m\ge 0$.
      \end{itemize}
\end{itemize}
\end{prop}

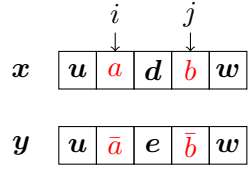
\begin{figure}[!t]
\centering
\begin{tikzpicture}
\draw (0,0) rectangle (0.5,0.5);
\draw (0.5,0) rectangle (1,0.5);
\draw (1,0) rectangle (1.5,0.5);
\draw (1.5,0) rectangle (2,0.5);
\draw (2,0) rectangle (2.5,0.5);
\path (0.25,0.25)node{$\bm{u}$}
      (0.75,0.25)node{\textcolor{red}{$a$}}
      (1.25,0.25)node{$\bm{d}$}
      (1.75,0.25)node{\textcolor{red}{$b$}}
      (2.25,0.25)node{$\bm{w}$}
      (-0.5,0.25)node{$\bm{x}$};
\draw (0,-1) rectangle (0.5,-0.5);
\draw (0.5,-1) rectangle (1,-0.5);
\draw (1,-1) rectangle (1.5,-0.5);
\draw (1.5,-1) rectangle (2,-0.5);
\draw (2,-1) rectangle (2.5,-0.5);
\path (0.25,-0.75)node{$\bm{u}$}
      (0.75,-0.75)node{\textcolor{red}{$\bar{a}$}}
      (1.25,-0.75)node{$\bm{e}$}
      (1.75,-0.75)node{\textcolor{red}{$\bar{b}$}}
      (2.25,-0.75)node{$\bm{w}$}
      (-0.5,-0.75)node{$\bm{y}$};
\draw[->](0.75,0.75)--(0.75,0.5);
\draw[->](1.75,0.75)--(1.75,0.5);
\draw (0.75,1)node{$i$};
\draw (1.75,1)node{$j$};
\end{tikzpicture}
\caption{We denote $i$ and $j$ to be the leftmost and rightmost indices where $\bm{x}$ and $\bm{y}$ differ.}
\label{fig_lemtypB}
\end{figure}

\begin{lem}\label{lem_typB}
  Suppose $\bm{x},\bm{y}\in\Sigma_2^n$. If $\left|I_{1}(\bm{x})\cap I_{1}(\bm{y})\right|=1$, then $\bm{x},\bm{y}$ are Type-B confusable. 
\end{lem}
\begin{IEEEproof}
Since $\left|I_{1}(\bm{x})\cap I_{1}(\bm{y})\right|=1$, then by \Cref{lem_typA},  $\bm{x}$ and $\bm{y}$ are not Type-A confusable, which further implies that $d_H(\bm{x},\bm{y})\ge 2$. So we can assume that
$$\bm{x}=\bm{u}a\bm{d}b\bm{w}, \text{ and }\bm{y}=\bm{u}\bar{a}\bm{e}\bar{b}\bm{w},
$$
where $a,b\in\Sigma_2$ and $\bm{u},\bm{d},\bm{e},\bm{w}\in\Sigma_2^{*}$. Let $i,j$ be the leftmost and rightmost indices, respectively, where $\bm{x}$ and $\bm{y}$ differ. See \Cref{fig_lemtypB} for a depiction.
\par
Suppose that $\bm{d}=\bm{e}=\emptyset$. If $a=\bar{b}$, then $\bm{x}$, $\bm{y}$ are Type-A confusable, which is a contradiction. If $a=b$, then $|wt_H(\bm{x})-wt_H(\bm{y})|=2$ and thus $\left|I_{1}(\bm{x})\cap I_{1}(\bm{y})\right|=0$, a contradiction. Therefore, we conclude that $\bm{d}$ and $\bm{e}$ are both nonempty.

Let $\bm{z}=z_1z_2\cdots z_{n+1}$ and $\{\bm{z}\}=I_{1}(\bm{x})\cap I_{1}(\bm{y})$.
Suppose that $z_k$ and $z_\ell$ are inserted in $\bm{x}$ and $\bm{y}$ to get $\bm{z}$, respectively. Then there are only two possibilities: $k\le i$ and $\ell>j$, or $k> j$ and $\ell\le i$.
If $k\le i$ and $\ell>j$, then $\bm{z}=\bm{u}\bar{a}a\bm{d}b\bm{w}=\bm{u}\bar{a}\bm{e}\bar{b}b\bm{w}$ and thus $a\bm{d}=\bm{e}\bar{b}$. This means that there exists $\bm{v}\in\Sigma_2^{*}$ such that $\bm{d}=\bm{v}\bar{b}$ and $\bm{e}=a\bm{v}$. Therefore, $\bm{x},\bm{y}$ are Type-B confusable.
If $k> j$ and $\ell\le i$, the proof is similar.
\end{IEEEproof}

\begin{figure}[!t]
\centering
\begin{tikzpicture}
\draw[fill=pink] (-0.2,0) circle [radius=1.2cm];
\draw (1.5,0) ellipse[x radius=2.4cm,y radius=1.5cm];
\draw (2,0) ellipse[x radius=3.8cm, y radius=1.8cm];
\draw (-0.7,0.25) node{\small{$I_1=2$}};
\draw (-0.45,-0.25) node{\small{$I_2=2n+4$}};
\draw (2.5,0) node{\small{$I_1=1$}};
\draw (2.3,-0.5) node{\small{$n+3\le I_2\le n+5$}};
\draw (4.7,0.25) node{$I_1=0$};
\draw (4.7,-0.25) node{$I_2\le 6$};
\draw[->,thick] (-1,1.5)--(-1,0.6);
\draw (-1,1.7)node{\footnotesize{Type-A}};
\draw[->,thick] (1,2)--(1.5,0.6);
\draw (1,2.2)node{\footnotesize{Type-B not Type-A}};
\draw[->,thick] (4.5,2)--(4.5,0.6);
\draw (4.5,2.2)node{\footnotesize{neither Type-A nor Type-B}};
\end{tikzpicture}
\caption{The relation of two kinds of confusability and the intersection sizes, where $I_1\triangleq\left|I_1\left(\bm{x}\right)\cap I_1\left(\bm{y}\right)\right|$ and $I_2\triangleq\left|I_2\left(\bm{x}\right)\cap I_2\left(\bm{y}\right)\right|$.}
\label{fig_typAandB}
\end{figure}
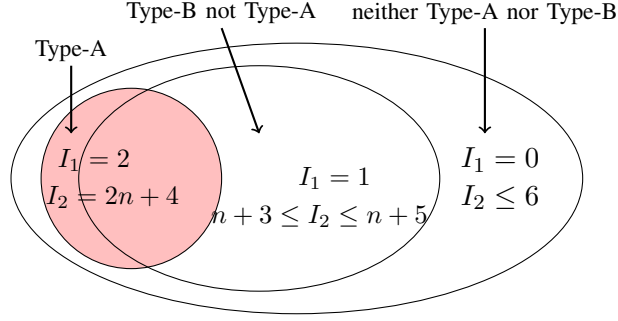

By \Cref{lem_typB}, if $\left|I_{1}(\bm{x})\cap I_{1}(\bm{y})\right|=1$, then we must have $\left\{\bm{x},\bm{y}\right\}=\left\{\bm{u}a\bar{a}\bm{v}b\bm{w},\bm{u}\bar{a}\bm{v}b\bar{b}\bm{w}\right\}$ for some $\bm{u}$, $\bm{v}$, $\bm{w}$ $\in\Sigma_2^{*}$ and $a,b\in\Sigma_2$. Consequently, $\bm{z}=\bm{u}a\bar{a}\bm{v}b\bar{b}\bm{w}$ is the unique sequence in $I_{1}(\bm{x})\cap I_{1}(\bm{y})$.  Combining \Cref{lem_typA} and \Cref{lem_typB}, we completely characterize the structures of two binary sequences for which the intersection of $1$-insertion balls is of a certain size. See the following corollary and \Cref{fig_typAandB}.
\begin{cor}\label{cor_summary}
  Let $\bm{x},\bm{y}\in\Sigma_2^n$ be distinct. Then
  \begin{itemize}
    \item $\left|I_{1}(\bm{x})\cap I_{1}(\bm{y})\right|=0$ if and only if $\bm{x}$ and $\bm{y}$ are neither Type-A confusable, nor Type-B confusable;
    \item $\left|I_{1}(\bm{x})\cap I_{1}(\bm{y})\right|=1$ if and only if $\bm{x}$ and $\bm{y}$ are Type-B confusable, but not Type-A confusable;
    \item $\left|I_{1}(\bm{x})\cap I_{1}(\bm{y})\right|=2$ if and only if $\bm{x}$ and $\bm{y}$ are Type-A confusable.
  \end{itemize}
\end{cor}

Now we proceed to estimate the intersection size of two $2$-insertions balls, that is  $\left|I_{2}(\bm{x})\cap I_{2}(\bm{y})\right|$, based on the value of $\left|I_{1}(\bm{x})\cap I_{1}(\bm{y})\right|$. By Equation~(\ref{Nl=1}), we have that $N^{+}(n,2)=2n+4$. That is, $\left| I_2(\bm{x})\cap I_2(\bm{y})\right|\leq 2n+4$ for any $\bm{x}\ne \bm{y}\in\Sigma_2^n$. The following lemma further restricts its value
in the case that $\left|I_{1}(\bm{x})\cap I_{1}(\bm{y})\right|=1$.

\begin{lem}\label{lem_basecase2}
Let $n\ge 3$ and $\bm{x},\bm{y}\in\Sigma_2^n$. If $|I_{1}(\bm{x})\cap I_{1}(\bm{y})|=1$, then $n+3\leq \left|I_{2}(\bm{x})\cap I_{2}(\bm{y})\right|\le n+5$. In particular, when $\bm{x}=a\bar{a}\bm{v}b$ and $\bm{y}=\bar{a}\bm{v}b\bar{b}$, we have
\begin{itemize}
 \item $\left|I_{2}(\bm{x})\cap I_{2}(\bm{y})\right|=n+3$ if and only if $a\bar{a}\bm{v}$ and $\bm{v}b\bar{b}$ are neither Type-B confusable, nor Type-A confusable;
 \item $\left|I_{2}(\bm{x})\cap I_{2}(\bm{y})\right|=n+4$ if and only if $a\bar{a}\bm{v}$ and $\bm{v}b\bar{b}$ are Type-B confusable, but not Type-A confusable;
 \item $\left|I_{2}(\bm{x})\cap I_{2}(\bm{y})\right|=n+5$ if and only if $a\bar{a}\bm{v}$ and $\bm{v}b\bar{b}$ are Type-A confusable.
\end{itemize}
\end{lem}
\begin{IEEEproof}
Since $\left|I_{1}(\bm{x})\cap I_{1}(\bm{y})\right|=1$, we conclude that $\bm{x},\bm{y}$ are Type-B confusable by \Cref{lem_typB}.
Assume that $\bm{x}=\bm{u}a\bar{a}\bm{v}b\bm{w}$ and $\bm{y}=\bm{u}\bar{a}\bm{v}b\bar{b}\bm{w}$ for some $a,b\in\Sigma_2$ and $\bm{u},\bm{v},\bm{w}\in\Sigma_2^{*}$. Let $I_{1}(\bm{x})\cap I_{1}(\bm{y})=\{\bm{z}\}$. Then $\bm{z}=\bm{u}a\bar{a}\bm{v}b\bar{b}\bm{w}$. Let $S=I_{2}(\bm{x})\cap I_{2}(\bm{y})$. 
We prove that $n+3\leq|S|\leq n+5$ by induction on the length $n$.

The base case is $\bm{u}=\bm{w}=\emptyset$, that is, $\bm{x}=a\bar{a}\bm{v}b,\bm{y}=\bar{a}\bm{v}b\bar{b}$ and $\bm{z}=a\bar{a}\bm{v}b\bar{b}$. In this case, we have $S=S^{a}\cup S_{\bar{b}}\cup S_{b}^{\bar{a}}$ and
$$
\begin{array}{l}
  S^{a}=a\left(I_2(\bar{a}\bm{v}b)\cap I_{1}(\bar{a}\bm{v}b\bar{b})\right)=aI_{1}(\bar{a}\bm{v}b\bar{b})\subseteq I_{1}(\bm{z}), \\
  S_{\bar{b}}=\left(I_{1}(a\bar{a}\bm{v}b)\cap I_2(\bar{a}\bm{v}b)\right)\bar{b}=I_{1}(a\bar{a}\bm{v}b)\bar{b}\subseteq I_{1}(\bm{z}), \\
  S_{b}^{\bar{a}}=\bar{a}\left(I_{1}(a\bar{a}\bm{v})\cap I_{1}(\bm{v}b\bar{b})\right)b.
\end{array}
$$
The second equality in the first line follows from the fact that $I_2(\bar{a}\bm{v}b)\cap I_{1}(\bar{a}\bm{v}b\bar{b})=I_{1}(\bar{a}\bm{v}b\bar{b})$, and similarly for the second equality in the second line.

By the form of $\bm{z}$, we have $I_{1}(\bm{z})\subseteq S^{a}\cup S_{\bar{b}}$ and $I_{1}(\bm{z})\cap S_{b}^{\bar{a}}=\emptyset$. Thus $S$ is the disjoint union of $I_{1}(\bm{z})$ and $S_{b}^{\bar{a}}$.
By \Cref{Nl=0}, $|I_{1}(\bm{z})|=n+3$. So it remains to show that $|S_{b}^{\bar{a}}|\leq 2$, or equivalently $a\bar{a}\bm{v}\neq \bm{v}b\bar{b}$ by \Cref{Nl=1}. Otherwise,
if $a\bar{a}\bm{v}=\bm{v}b\bar{b}$, then $a=b$ and $\bm{v}=(a\bar{a})^m$, or $a=\bar{b}$ and $\bm{v}=(a\bar{a})^ma$ for some $m\ge 0$. For both cases, $\bm{x}$ and $\bm{y}$ are Type-A confusable by \Cref{prop_summary} (2), which contradicts the fact that $\left|I_{1}(\bm{x})\cap I_{1}(\bm{y})\right|=1$. This completes the proof of the base case by \Cref{cor_summary}.

In the base case, we have shown that $S=I_1(\bm{z})\cup J({\bm{x}},{\bm{y}})$  for some set $J({\bm{x}},{\bm{y}})$ (i.e. $S_{b}^{\bar{a}}$) of size at most two.  Suppose  we have proved the same conclusion for the codeword length $n-1$, and we will prove it for length $n$. Now $\bm{x}=\bm{u}a\bar{a}\bm{v}b\bm{w}$, $\bm{y}=\bm{u}\bar{a}\bm{v}b\bar{b}\bm{w}$ and $\bm{z}=\bm{u}a\bar{a}\bm{v}b\bar{b}\bm{w}$.
Without loss of generality, we assume that $|\bm{u}|\ge 1$ and $\bm{u}=c\bm{u}^{*}$ for some $c\in\Sigma_2$ and $\bm{u}^{*}\in\Sigma_2^{*}$. Then $S^{\bar{c}}$ $=$ $\bar{c}\left(I_1(\bm{x})\cap I_1(\bm{y})\right)$ $=$ $\{\bar{c}\bm{z}\}=I_1(\bm{z})^{\bar{c}}$. Denote $\widetilde{\bm{x}}=\bm{u}^{*}a\bar{a}\bm{v}b\bm{w}$, $\widetilde{\bm{y}}=\bm{u}^{*}\bar{a}\bm{v}b\bar{b}\bm{w}$ and $\widetilde{\bm{z}}=\bm{u}^{*}a\bar{a}\bm{v}b\bar{b}\bm{w}$, which are obtained from $\bm{x},\bm{y},\bm{z}$ by deleting the first element $c$.
By the induction hypothesis, $I_2(\widetilde{\bm{x}})\cap I_2(\widetilde{\bm{y}})=I_1(\widetilde{\bm{z}})\cup J(\widetilde{\bm{x}},\widetilde{\bm{y}})$, for some set $J(\widetilde{\bm{x}},\widetilde{\bm{y}})$ of size at most two.
Then $S^c=c\left(I_2(\widetilde{\bm{x}})\cap I_2(\widetilde{\bm{y}})\right)=cI_1(\widetilde{\bm{z}})\cup cJ(\widetilde{\bm{x}},\widetilde{\bm{y}})=I_1(\bm{z})^c\cup cJ(\widetilde{\bm{x}},\widetilde{\bm{y}})$. Hence $S=S^{\bar{c}}\cup S^c=I_1(\bm{z})\cup cJ(\widetilde{\bm{x}},\widetilde{\bm{y}})$, and this completes the proof.
\end{IEEEproof}

By \Cref{lem_basecase2}, we are able to give a rough estimate of $\left| I_2(\bm{x})\cap I_2(\bm{y})\right|$ based on the different values of $\left| I_1(\bm{x})\cap I_1(\bm{y})\right|$, and consequently based on the confusability of $\bm{x}$ and $\bm{y}$ by \Cref{cor_summary}. See the following lemma and \Cref{fig_typAandB}.

\begin{lem}\label{lem_max2}
  Let $\bm{x}\ne \bm{y}\in\Sigma_2^n$ where $n\ge 4$. Then
   \begin{enumerate}[(i)]
     \item $\left| I_1(\bm{x})\cap I_1(\bm{y})\right|=2$ if and only if $\left| I_2(\bm{x})\cap I_2(\bm{y})\right|=2n+4$;
     \item $\left| I_1(\bm{x})\cap I_1(\bm{y})\right|=1$ if and only if $n+3\leq \left| I_2(\bm{x})\cap I_2(\bm{y})\right|\leq n+5$;
     \item $\left| I_1(\bm{x})\cap I_1(\bm{y})\right|=0$ if and only if $ \left| I_2(\bm{x})\cap I_2(\bm{y})\right|\leq 6$.
   \end{enumerate}
\end{lem}
\begin{IEEEproof}
(i) We prove the sufficiency by contradiction. If $\left| I_1(\bm{x})\cap I_1(\bm{y})\right|=0$, that is $d_L(\bm{x},\bm{y})\geq 2$, then $\left| I_2(\bm{x})\cap I_2(\bm{y})\right|\le N^{+}(n,2,2)= 6<2n+4$ by \Cref{comN}. If $\left| I_1(\bm{x})\cap I_1(\bm{y})\right|=1$, then $\left| I_2(\bm{x})\cap I_2(\bm{y})\right|\le n+5<2n+4$ by \Cref{lem_basecase2}. So we must have $\left| I_1(\bm{x})\cap I_1(\bm{y})\right|=2$.

For the necessity, recall that $\left| I_2(\bm{x})\cap I_2(\bm{y})\right|\le 2n+4$.  Since $\left| I_1(\bm{x})\cap I_1(\bm{y})\right|$ $=2$, by \Cref{lem_typA}, there must be some $\bm{u},\bm{w}\in\Sigma_2^{*}$, $a\in\Sigma_2$ and $m\ge 0$, such that
$$
\left\{
\begin{array}{l}
\bm{x}=\bm{u}(a\bar{a})^{m+1}\bm{w}\\
\bm{y}=\bm{u}(\bar{a}a)^{m+1}\bm{w}
\end{array}
\right.
\text{\ or\ }
\left\{
\begin{array}{l}
\bm{x}=\bm{u}(a\bar{a})^{m}a\bm{w}\\
\bm{y}=\bm{u}(\bar{a}a)^{m}\bar{a}\bm{w}
\end{array}.
\right.
$$
For the left case, let $\bm{z}_1=\bm{u}(\bar{a}a)^{m+1}\bar{a}\bm{w}$ and $\bm{z}_2=\bm{u}(a\bar{a})^{m+1}a\bm{w}$. Then  $I_1(\bm{x})\cap I_1(\bm{y})=\{\bm{z}_1,\bm{z}_2\}$ and $I_1(\bm{z}_1)\cap I_1(\bm{z}_2)=\{\bm{u}(\bar{a}a)^{m+2}\bm{w},\bm{u}(a\bar{a})^{m+2}\bm{w}\}$. Noting that
$\left|I_1(\bm{z}_1)\right|=\left|I_1(\bm{z}_2)\right|=n+3$ by \Cref{Nl=0}, we have $\left| I_2(\bm{x})\cap I_2(\bm{y})\right|\ge|I_1(\bm{z}_1)\cup I_1(\bm{z}_2)|$ $\ge 2n+4$. Therefore, $\left| I_2(\bm{x})\cap I_2(\bm{y})\right|= 2n+4$.
For the right case, the proof is similar.

The claim (ii) is clear by \Cref{lem_basecase2}, the claim (i) and the fact that $\left| I_2(\bm{x})\cap I_2(\bm{y})\right|\le 6$ if $\left| I_1(\bm{x})\cap I_1(\bm{y})\right|=0$.
The claim (iii) then follows too.
\end{IEEEproof}

\begin{figure}[!t]
\centering
\begin{tikzpicture}
\draw[fill=red!30,opacity=0.3] (-2,-0.85) rectangle (2,3);
\draw (0,2.75) node{\tiny{$N_1\ge 3, N_2\ge 2n+5$}};
\draw[fill=blue!20,opacity=0.5] (-1.6,-0.75) rectangle (1.6,2.5);
\draw (0,2.26) node{\tiny{$N_1=2$}};
\draw (0,2.05) node{\tiny{$n+5< N_2\le 2n+4$}};
\draw (-1.3,-0.65) rectangle (1.3,1.8);
\draw (0,1.55) node{\tiny{$ N_2= n+4,n+5$}};
\draw[fill=yellow,opacity=0.7] (-1,-0.55) rectangle (1,1.3);
\draw (0,0.95) node{\tiny{$N_1=1$}};
\draw (0,0.75) node{\tiny{$6< N_2\le n+3$}};
\draw (-0.7,-0.5) rectangle (0.7,0.5);
\draw (0,0) node{\tiny{$1\le N_2\le 6$}};
\end{tikzpicture}
\caption{Relations between $(n,N_1;B^{I(1)})$-reconstruction codes and $(n,N_2;B^{I(2)})$-reconstruction codes, and a hierarchy relation for different values of $N$. Values of $N_1$ and $N_2$ in the same layer satisfy the ``if and only if'' statement, that is an $(n,N_1;B^{I(1)})$-reconstruction code is an $(n,N_2;B^{I(2)})$-reconstruction code, and vice verse. Values $N$ in a smaller square and values $N'$ in a bigger square satisfy the ``inclusion'' statement,  that is, an $(n,N;B^{I(t)})$-reconstruction code is an $(n,N';B^{I(t)})$-reconstruction code for $t=1,2$, but the converse is not true.}
\label{fig_inclusionrelationship}
\end{figure}
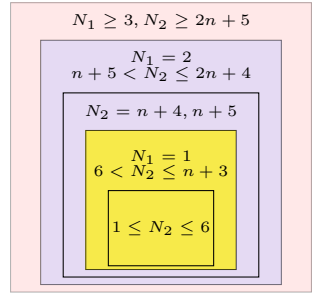

By \Cref{lem_max2} and \Cref{fig_typAandB}, it is immediate to have the following
 relations between reconstruction codes for two insertions and those for single insertion under certain conditions. See \Cref{fig_inclusionrelationship} for a depiction.

\begin{cor}\label{cor_n5} Let $n\geq 4$ and $\mathcal{C} \subset \Sigma_2^n$.
\begin{enumerate}[(i)]
  \item When $n+5<N\le 2n+4$, $\mathcal{C}$ is an $(n,N;B^{I(2)})$-reconstruction code if and only if $\mathcal{C}$ is an $(n,2;B^{I(1)})$-reconstruction code.
  \item When $6<N\le n+3$, $\mathcal{C}$ is an $(n,N;B^{I(2)})$-reconstruction code if and only if $\mathcal{C}$ is an $(n,1;B^{I(1)})$-reconstruction code.
\end{enumerate}
\end{cor}

By now, we are able to determine $\rho(n,N;B^{I(2)})$ for most values of $N$. First, since $\left| I_2(\bm{x})\cap I_2(\bm{y})\right|\leq 2n+4$ for any $\bm{x}\ne \bm{y}\in\Sigma_2^n$, we have $\nu_2 (\Sigma_2^n)=2n+4$ and $\rho(n,N;B^{I(2)})=0$ for any $N>2n+4$.
By \Cref{lem_max2}, we  see that if $N=n+4,n+5$, an $(n,N;B^{I(2)})$-reconstruction code must be an $(n,2;B^{I(1)})$-reconstruction code, and thus $\rho(n,N;B^{I(2)})\ge\log_2\log_2(n)+\Omega(1)$ by \Cref{thm_1inser}; if $2\le N\le 6$, an $(n,N;B^{I(2)})$-reconstruction code must be an $(n,1;B^{I(1)})$-reconstruction code, 
and thus $\rho(n,N;B^{I(2)})\ge\log_2(n)+\Omega(1)$  by \Cref{thm_1inser}. The last case  $N=1$, corresponds to a $2$-insertion error-correcting code, or equivalently a $2$-deletion error-correcting code, whose best known upper bound on the optimal redundancy is $4\log_2(n)+O(\log_2\log_2(n))$ \cite[Theorem 1]{wz4}, and the best known lower bound is $2\log_2(n)-O(1)$ (see \cite{wz9}). \Cref{prop_summary1} summarizes the information given above.

\begin{prop}\label{prop_summary1}
\begin{equation*}
\rho(n,N;B^{I(2)})=
\begin{cases}
  0, & \mbox{if } N>2n+4, \\
  \log_2\log_2(n)+\Theta(1), & \mbox{if } n+5<N\le 2n+4, \\
  \ge\log_2\log_2(n)+\Omega(1), & \mbox{if } N=n+4,n+5, \\
  \log_2(n)+\Theta(1), & \mbox{if } 6<N\le n+3, \\
  \Theta(\log_2(n)), & \mbox{if } 1\le N\le 6.
\end{cases}
\end{equation*}
\end{prop}

Therefore, the remaining cases are $N\in\{n+4,n+5\}$ and $2\le N\le 6$,  which  will be handled in the rest of this paper. In fact, we construct asymptotically optimal reconstruction codes for $N\in\{n+4,n+5\}$ with redundancy $\log_2\log_2(n)+O(1)$,  and for $N\in\{5,6\}$ with redundancy $\log_2(n)+14\log_2\log_2(n)+O(1)$, thus determine the asymptotically optimal redundancy for all $N\geq 5$.

\section{Reconstruction Codes for $N\in\{n+4,n+5\}$}\label{sec_np4np5}
In this section, we construct reconstruction codes for $N\in\{n+4,n+5\}$ with redundancy as small as possible. The main idea is as follows. First, we choose an $(n,n+6;B^{I(2)})$-reconstruction code with small redundancy, or equivalently an $(n,2;B^{I(1)})$-reconstruction code $\mathcal{C}_1$ (see \Cref{cor_n5} (i)), which can be found in \cite[Theorem 17]{wz2}. Then we impose additional constraints on $\mathcal{C}_1$ such that $\left|I_2(\bm{x})\cap I_2(\bm{y})\right|\notin\{n+4, n+5\}$ under these constraints, and consequently we obtain  an $(n,n+4;B^{I(2)})$-reconstruction code $\mathcal{C}$.


The main task in our construction is to characterize the structures of two sequences $\bm{x},\bm{y}$ when $\left| I_2(\bm{x})\cap I_2(\bm{y})\right|=n+4$ or $n+5$. By \Cref{lem_max2}, we have $\left| I_1(\bm{x})\cap I_1(\bm{y})\right|=1$. Then by \Cref{lem_typB}, we can assume that
$$
  \bm{x}=\bm{u}a\bar{a}\bm{v}b\bm{w} \text{ and }
  \bm{y}=\bm{u}\bar{a}\bm{v}b\bar{b}\bm{w}
$$
for some $\bm{u},\bm{v},\bm{w}\in\Sigma_2^{*}$ and $a,b\in\Sigma_2$, where $\bm{v}$ satisfies neither of the following conditions by  \Cref{prop_summary}:
\begin{equation}\label{eq_rsv}
\begin{array}{c}
a=b\text{ and }\bm{v}=(a\bar{a})^m \text{ for some } m\ge 0,\\
a=\bar{b}\text{ and }\bm{v}=(a\bar{a})^ma  \text{ for some } m\ge 0.
\end{array}
\end{equation}
Let $\bm{z}=\bm{u}a\bar{a}\bm{v}b\bar{b}\bm{w}$. In the proof of \Cref{lem_basecase2}, we have shown that  $I_2(\bm{x})\cap I_2(\bm{y})=I_1(\bm{z})\cup J({\bm{x}},{\bm{y}})$  for some set $J({\bm{x}},{\bm{y}})$ of size at most two. The following result reveals more information about $J({\bm{x}},{\bm{y}})$,
whose proof is deferred to Appendix \ref{appendix1}.
\begin{lem}\label{lem_reduced}
  Let $\bm{x}=\bm{u}a\bar{a}\bm{v}b\bm{w}$ and $\bm{y}=\bm{u}\bar{a}\bm{v}b\bar{b}\bm{w}$ for some $\bm{u},\bm{v},\bm{w}$ $\in\Sigma_2^{*}$ and $a,b\in\Sigma_2$, and let $\bm{z}=\bm{u}a\bar{a}\bm{v}b\bar{b}\bm{w}$. If $\bm{x}$ and $\bm{y}$ are not Type-A confusable, then
$$
\begin{array}{l}
  I_2(\bm{x})\cap I_2(\bm{y})
  =I_1(\bm{z})\cup \bm{u}\left(I_2(a\bar{a}\bm{v}b)\cap I_2(\bar{a}\bm{v}b\bar{b})\setminus I_1(a\bar{a}\bm{v}b\bar{b})\right)\bm{w},
\end{array}
$$
and the union is disjoint.
\end{lem}

Let $n^{\prime}$ be the length of $a\bar{a}\bm{v}b$. Then by \Cref{Nl=0}, we have $\left|I_1(\bm{z})\right|=n+3$ and $\left|I_1(a\bar{a}\bm{v}b\bar{b})\right|=n^{\prime}+3$. \Cref{lem_reduced} implies
\begin{align*}
\left| I_2(\bm{x})\cap I_2(\bm{y})\right|&=\left|I_1(\bm{z})\right|+\left|I_2(a\bar{a}\bm{v}b)\cap I_2(\bar{a}\bm{v}b\bar{b})\setminus I_1(a\bar{a}\bm{v}b\bar{b})\right|\\
&=n+3+\left(\left|I_2(a\bar{a}\bm{v}b)\cap I_2(\bar{a}\bm{v}b\bar{b})\right|-n^{\prime}-3\right).
\end{align*}
By \Cref{eq_rsv} and \Cref{prop_summary},  $a\bar{a}\bm{v}b$ and $\bar{a}\bm{v}b\bar{b}$ are Type-B confusable but not Type-A confusable, and hence $\left|I_1(a\bar{a}\bm{v}b)\cap I_1(\bar{a}\bm{v}b\bar{b})\right|=1$ by \Cref{cor_summary}. Then $\left|I_2(a\bar{a}\bm{v}b)\cap I_2(\bar{a}\bm{v}b\bar{b})\right|\in\{n^{\prime}+3,n^{\prime}+4, n^{\prime}+5\}$ by \Cref{lem_basecase2}. We also have $\left|I_2(\bm{x})\cap I_2(\bm{y})\right|\in\{n+3,n+4, n+5\}$ due to the fact $\left| I_1(\bm{x})\cap I_1(\bm{y})\right|=1$.
So the above equality implies that $\left| I_2(\bm{x})\cap I_2(\bm{y})\right|=n+4$ or $n+5$ if and only if $\left| I_2(a\bar{a}\bm{v}b)\cap I_2(\bar{a}\bm{v}b\bar{b})\right|=n^{\prime}+4$ or $n^{\prime}+5$, respectively.  Thus, we can further assume that
$
  \bm{x}=a\bar{a}\bm{v}b \text{ and }
  \bm{y}=\bar{a}\bm{v}b\bar{b}.
$ The following theorem characterizes the conditions when $\left| I_2(a\bar{a}\bm{v}b)\cap I_2(\bar{a}\bm{v}b\bar{b})\right|=n^{\prime}+4$ or $n^{\prime}+5$.

\begin{table}[!htbp]
  \centering
  \caption{case (\Rnum{2}) of \Cref{thm_classification1}}\label{tb_thmC1}
  \begin{tabular}{c|c|c|c}
    \hline
     & $\bm{v}$ & Conditions&Row Number \\
    \hline
    \multirow{4}{*}{$a=b$}&$(a\bar{a})^i\bar{a}^j(a\bar{a})^{k}$&$i\ge 0,j\ge 2,k\ge 0$&1\\
    \cline{2-4}
    &$(a\bar{a})^ia^j(a\bar{a})^{k}$&$i\ge 0,j\ge 2,k\ge 0$&2\\
    \cline{2-4}
    & $(a\bar{a})^i(\bar{a}a\bar{a})^j(\bar{a}a)^{k}\bar{a}$& $i\ge 0,j\ge 1,k\ge 0$&3\\
    \cline{2-4}
    & $(a\bar{a})^ia(a\bar{a}a)^j(a\bar{a})^{k}$& $i\ge 0,j\ge 1,k\ge 0$&4\\
    \hline
   \multirow{4}{*}{$a=\bar{b}$}&$(a\bar{a})^i\bar{a}^j(\bar{a}a)^{k}$&$i\ge 0,j\ge 1,k\ge 0$&5\\
   \cline{2-4}
   &$(a\bar{a})^ia^j(\bar{a}a)^{k}$& $i\ge 0,j\ge 3,k\ge 0$&6\\
   \cline{2-4}
   &$(a\bar{a})^i(\bar{a}a\bar{a})^j(\bar{a}a)^{k}$&$i\ge 0,j\ge 1,k\ge 0$&7\\
   \cline{2-4}
   &$(a\bar{a})^ia(a\bar{a}a)^j(a\bar{a})^{k}a$ &$i\ge 0,j\ge 1,k\ge 0$&8\\
   \hline
  \end{tabular}
\end{table}

\begin{thm}\label{thm_classification1} Let $n^{\prime}\geq 3$.
Suppose that $a,b\in\Sigma_2$, and $\bm{v}\in \Sigma_2^{n^{\prime}-3}$ do not satisfy the conditions in (\ref{eq_rsv}). Then $\bm{x}=a\bar{a}\bm{v}b$ and $\bm{y}=\bar{a}\bm{v}b\bar{b}$ satisfy the following properties.
  \begin{enumerate}[(i)]
    \item $\left| I_2(\bm{x})\cap I_2(\bm{y})\right|=n^{\prime}+5$ if and only if
    \begin{itemize}
      \item  $a=b$ and $\bm{v}\in\{(a\bar{a})^{i}a(a\bar{a})^{j},(a\bar{a})^{i}(\bar{a}a)^{j}\bar{a}\}$ for some $i,j\ge 0$; or
      \item  $a=\bar{b}$ and $\bm{v}\in\{(a\bar{a})^{i}(\bar{a}a)^{j},(a\bar{a})^{i}aa(\bar{a}a)^{j}\}$ for some $i,j\ge 0$.
    \end{itemize}
    \item $\left| I_2(\bm{x})\cap I_2(\bm{y})\right|=n^{\prime}+4$ if and only if $a,b$ and $\bm{v}$ satisfy one of the conditions listed in \Cref{tb_thmC1}.
  \end{enumerate}
\end{thm}
\begin{IEEEproof} Denote $l=n^{\prime}-3$. Let
 $\widetilde{\bm{x}}= a\bar{a}\bm{v}=\tilde{x}_1\cdots \tilde{x}_{l+2}$ and $\widetilde{\bm{y}}= \bm{v}b\bar{b}=\tilde{y}_1\cdots \tilde{y}_{l+2}$. Then
 \begin{equation}\label{eq_1}
  \left\{
  \begin{array}{l}
      \tilde{x}_1=a, \tilde{x}_2=\bar{a};\\
       \tilde{x}_{i+2}=\tilde{y}_i, \text{ for all }i=1,\ldots,l;\\
     \tilde{y}_{l+1}=b, \tilde{y}_{l+2}=\bar{b}.
  \end{array}
  \right.
\end{equation}

(\Rnum{1}). From \Cref{lem_basecase2}, we know that $\left| I_2(\bm{x})\cap I_2(\bm{y})\right|=n^{\prime}+5$ if and only if $\widetilde{\bm{x}}$ and $\widetilde{\bm{y}}$ are Type-A confusable. That is to say, $\widetilde{\bm{x}}=\bm{u}\bm{c}\bm{w}$ and $\widetilde{\bm{y}}=\bm{u}\bar{\bm{c}}\bm{w}$
  for some $\bm{u},\bm{c},\bm{w}\in\Sigma_2^{*}$, where $\bm{c}$ is an alternating sequence of length at least one. Denote $s=|\bm{u}|\geq 0$ and $t= |\bm{c}|\geq 1$. Then

  \begin{equation}\label{eq_2}
  \left\{
  \begin{array}{ll}
      \tilde{x}_i=\tilde{y}_i, \text{ for }i=1,\ldots,s,s+t+1,\ldots,l+2 &\text{ by $\bm{u}$ and $\bm{w}$};\\
       \tilde{x}_{i}=\overline{\tilde{y}_i}, \text{ for }i=s+1,\ldots,s+t &\text{ by $\bm{c}$ and $\bar{\bm{c}}$};\\
     \tilde{x}_i=\tilde{x}_{i+2}, \text{ for }i=s+1,\ldots,s+t-2 &\text{ by alternating of $\bm{c}$}.
  \end{array}
  \right.
\end{equation}
  Since $wt_H(\widetilde{\bm{x}})=wt_H(\widetilde{\bm{y}})=wt_H(\bm{v})+1$, we have $|\bm{c}|=d_H(\widetilde{\bm{x}},\widetilde{\bm{y}})\ge 2$.
   If $|\bm{c}|\geq 3$, consider the prefix $\tilde{x}_{s+1}\tilde{x}_{s+2}\tilde{x}_{s+3}$ of $\bm{c}$. By the third line of \Cref{eq_2} and the second line of \Cref{eq_1}, we have $\tilde{x}_{s+1}=\tilde{x}_{s+3}$ and $\tilde{x}_{s+3}=\tilde{y}_{s+1}$, respectively. Thus $\tilde{x}_{s+1}=\tilde{y}_{s+1}$, which contradicts the second line of \Cref{eq_2}.
 So we have  $|\bm{c}|=2$, and \Cref{eq_2} becomes
 \begin{equation}\label{eq_3}
  \left\{
  \begin{array}{l}
      \tilde{x}_i=\tilde{y}_i, \text{ for }i=1,\ldots,s,s+3,\ldots,l+2;\\
       \tilde{x}_{i}=\overline{\tilde{y}_i}, \text{ for }i=s+1,s+2;\\
     \tilde{x}_{s+1}=\overline{\tilde{x}_{s+2}}.
  \end{array}
  \right.
\end{equation}

Next, we determine the explicit structure of $\bm{v}$ according to the values of $s$ and $l$. Write $\bm{v}=v_1\cdots v_{l}$. Then
\begin{equation}\label{eq_4}
v_i=\tilde{x}_{i+2}=\tilde{y}_i, \text{ for all }i=1,\ldots,l.
\end{equation}
For convenience,  let $v_{-1}=\tilde{x}_1=a$,  $v_0=\tilde{x}_2=\bar{a}$. Since $l\ge s$, the first two lines of \Cref{eq_1} and the first line of \Cref{eq_3} imply that $v_{-1}v_0\cdots v_s$ is alternating. Therefore, 
\begin{equation}\label{eq_alterv}
  v_{-1}v_0\cdots v_s=
  \left\{
  \begin{array}{ll}
   (a\bar{a})^{\frac{s+2}{2}} &  \text{ if }s\text{ is even}, \\
    (a\bar{a})^{\frac{s+1}{2}}a &  \text{ if }s\text{ is odd}.
  \end{array}
  \right.
\end{equation}
We divide our discussion into three cases.
\begin{itemize}
\item  $l=s$. In this case, we have $v_l=\tilde{x}_{l+2}=\overline{\tilde{y}_{l+2}}=b$ and $v_{l-1}=\tilde{x}_{l+1}=\overline{\tilde{y}_{l+1}}=\bar{b}$. Therefore, $\bm{v}=(a\bar{a})^{\frac{l}{2}}$ and $a=\bar{b}$ when $l$ is even, or $\bm{v}=(a\bar{a})^{\frac{l-1}{2}}a$ and $a=b$ when $l$ is odd.
\item  $l=s+1$. In this case, we have $v_{l}=\tilde{x}_{l+2}=\tilde{y}_{l+2}=\bar{b}$ and $v_{l-1}=\tilde{x}_{l+1}=\overline{\tilde{y}_{l+1}}=\bar{b}$. Therefore, $\bm{v}=(a\bar{a})^{\frac{l-1}{2}}\bar{a}$ and $a=b$ when $l$ is odd, or $\bm{v}=(a\bar{a})^{\frac{l-2}{2}}aa$ and $a=\bar{b}$ when $l$ is even.

\item  $l\ge s+2$. In this case, $v_{i}=\tilde{y}_{i}=\overline{\tilde{x}_{i}}$ for $i=s+1,s+2$ by the second line of  \Cref{eq_3} and \Cref{eq_4}. So
  $v_{s+1}=\overline{v_{s+2}}$ by  the third line of  \Cref{eq_3}.
  Hence $v_{s+1}v_{s+2}\cdots v_{l}$ is alternating by the first line of \Cref{eq_3} and \Cref{eq_4}.
   Since  $l+2\ge s+4$, we have  $v_{l-1}=\tilde{x}_{l+1}=\tilde{y}_{l+1}=b$ and $v_{l}=\tilde{x}_{l+2}=\tilde{y}_{l+2}=\bar{b}$ by the first line of \Cref{eq_3}.  Now, combining with \Cref{eq_alterv}, we come to the desired forms of $\bm{v}$: for all $0\le s\le l-2$,
\begin{equation*}
\text{ when }a=\bar{b},\text{ }\bm{v}=v_1\cdots v_l=
  \left\{
  \begin{array}{ll}
   (a\bar{a})^{\frac{s}{2}}(\bar{a}a)^{\frac{l-s}{2}} &  \text{ if }l,s\text{ are both even}, \\
    (a\bar{a})^{\frac{s-1}{2}}aa(\bar{a}a)^{\frac{l-1-s}{2}} &  \text{ if }s\text{ is odd and }l\text{ is even},
  \end{array}
  \right.
\end{equation*}
and
\begin{equation*}
\text{ when }a=b,\text{ }\bm{v}=v_1\cdots v_l=
  \left\{
  \begin{array}{ll}
   (a\bar{a})^{\frac{s}{2}}(\bar{a}a)^{\frac{l-1-s}{2}}\bar{a} &  \text{ if }s\text{ is even and }l\text{ is odd}, \\
    (a\bar{a})^{\frac{s-1}{2}}a(a\bar{a})^{\frac{l-s}{2}} &  \text{ if }l,s\text{ are both odd}.
  \end{array}
  \right.
\end{equation*}
\end{itemize}
This completes the proof of (\Rnum{1}).
\par
(\Rnum{2}). The proof of this case is long and tedious, thus we move it to Appendix \ref{appendix2}.
\end{IEEEproof}

\Cref{thm_classification1} gives a full characterization of the structures of two sequences $\bm{x},\bm{y}\in \Sigma_2^n$ when $\left| I_2(\bm{x})\cap I_2(\bm{y})\right|=n+4$ or $n+5$, respectively. This will help to exclude such pairs of sequences from an  $(n,n+6;B^{I(2)})$-reconstruction code (or an  $(n,2;B^{I(1)})$-reconstruction code) to obtain an $(n,n+4;B^{I(2)})$-reconstruction code, as mentioned in the beginning of this section. We use the $(n,2;B^{I(1)})$-reconstruction code  in \cite[Theorem 17]{wz2} as a candidate for our construction. For that, we need the following notation.

%
  For any $\bm{x}\in\Sigma_2^n$, define
  $$
  \Inv{\bm{x}}=\left|\{(i,j):1\le i<j\le n\text{ and }x_i>x_j\}\right|.
  $$
By definition, for any $a\in\Sigma_2$ and any $\bm{u},\bm{v},\bm{w}\in\Sigma_2^{*}$,
\begin{equation}\label{eq_inv}
  |\Inv {\bm{u}a\bar{a}\bm{v}\bar{a}\bm{w}}-\Inv{\bm{u}\bar{a}\bm{v}\bar{a}a\bm{w}}|=|\Inv {a\bar{a}\bm{v}\bar{a}}-\Inv{\bar{a}\bm{v}\bar{a}a}|=N_{\bm{v}}(\bar{a})+2,
\end{equation}
where $N_{\bm{v}}(\bar{a})$ denotes the number of $\bar{a}$ appearing in $\bm{v}$. Let $R(n,\ell,t)$ denote the set of all sequences $\bm{x}\in\Sigma_2^n$ such that the length of any subword with period $\ell^{\prime}$ of $\bm{x}$ is at most $t$, for any $\ell^{\prime}\le \ell$. By definition, $R(n,\ell,t)\subseteq R(n,\jmath,s)$ if $\ell\geq \jmath$ and $t\leq s$.

\begin{lem}\label{lem_zq}
\textup{(\!\!\cite[Theorem 13]{wz3})}
  For $\ell\ge 1$, if $t\ge \lceil \log_2(n)\rceil +\ell+1$, we have that $|R(n,\ell,t)|\ge 2^{n-1}$.
\end{lem}


 For $n,P>0$, let $c\in\mathbb{Z}_{1+P}$ and $d\in\mathbb{Z}_2$. In \cite[Theorem 17]{wz2}, the authors defined the following code
\[\mathcal{C}_1=\mathcal{C}_1(n;c,d)=\{\bm{x}\in R(n,2,2P): \Inv{\bm{x}}\equiv c\pmod{1+P}, wt_H(\bm{x})\equiv d\pmod{2}\}.\]
They proved that $\mathcal{C}_1$ is an $(n,2;B^{I(1)})$-reconstruction code in the following way. The constraint on $wt_H(\bm{x})$ implies that the Hamming weights of any two distinct codewords $\bm{x}$ and $\bm{y}$ in $\mathcal{C}_1$ have the same parity and thus $d_H(\bm{x},\bm{y})\ge 2$. Therefore, if $\left|I_1(\bm{x})\cap I_1(\bm{y})\right|=2$, we can conclude that $\bm{x}=\bm{u}\bm{c}\bm{w}$ and $\bm{y}=\bm{u}\overline{\bm{c}}\bm{w}$ for some $\bm{u},\bm{w}\in\Sigma_2^{*}$ and some alternating sequence $\bm{c}$ of positive even length. On the one hand, it can be shown that $\left|\Inv{\bm{x}}-\Inv{\bm{y}}\right|=\frac{|\bm{c}|}{2}$. So the constraint on $\Inv{\bm{x}}$ leads to $(1+P)\mid \frac{|\bm{c}|}{2}$ and hence $1+P\le \frac{|\bm{c}|}{2}$. On the other hand, the length of $\bm{c}$ is upper bounded by $2P$ since $\bm{c}$ has period $2$. So we have $1+P\le \frac{|\bm{c}|}{2}\le P$, which is impossible. Therefore, $\mathcal{C}_1$ is an $(n,2;B^{I(1)})$-reconstruction code.

By \Cref{cor_n5} (i),
the code $\mathcal{C}_1$ is an $(n,n+6;B^{I(2)})$-reconstruction code. To obtain an $(n,n+4;B^{I(2)})$-reconstruction code from $\mathcal{C}_1$, we have to exclude pairs of sequences $\bm{x},\bm{y}\in \mathcal{C}_1$ satisfying that $\left|I_2(\bm{x})\cap I_2(\bm{y})\right|=n+4$ or $n+5$. By \Cref{thm_classification1}, such pairs have a common subword $\bm{v}$ with a certain periodic property.  Thus we can further bound the length of $\bm{v}$ and use the constraint on $\Inv {\bm{x}}$ to get a contradiction and exclude such pairs. See our construction below.



\begin{thm}\label{thm_construction1}\textup{($N=n+4$)}
  For integers $n\ge 4,P\ge 6$ such that $3\mid P$, let $c\in\mathbb{Z}_{1+P}$ and $d\in\mathbb{Z}_2$. Let
   \[\mathcal{C}(n;c,d)=\{\bm{x}\in R(n,3,\frac{P}{3}):  \Inv{\bm{x}}\equiv c\pmod{1+P}, wt_H(\bm{x})\equiv d\pmod{2}\}\}.\]
Then $\mathcal{C}(n;c,d)$ is an $(n,n+4;B^{I(2)})$-reconstruction code.  Furthermore, if $P=3\lceil \log_2(n)\rceil +12$, then $\mathcal{C}(n;c,d)$ has redundancy at most $\log_2(P+1)+2= \log_2\log_2(n)+O(1)$ for some choice of $c$ and $d$.

\end{thm}
\begin{IEEEproof}
Since $R(n,3,\frac{P}{3})\subseteq R(n,2,2P)$, that is $\mathcal{C}(n;c,d)\subseteq \mathcal{C}_1(n;c,d)$, it suffices to show that $\left|I_2(\bm{x})\cap I_2(\bm{y})\right|\notin \{n+4,n+5\}$ for any two distinct sequences $\bm{x},\bm{y}\in \mathcal{C}(n;c,d)$.

\par
If, on the contrary, there exists such a pair  $\bm{x},\bm{y}$, then without loss of generality, we can assume that
$$
  \bm{x}=\bm{u}a\bar{a}\bm{v}b\bm{w} \text{ and }
  \bm{y}=\bm{u}\bar{a}\bm{v}b\bar{b}\bm{w}
$$for some $\bm{u},\bm{v},\bm{w}$ $\in\Sigma_2^{*}$ and $a,b\in\Sigma_2$.
Since $wt_H(\bm{x})\equiv wt_H(\bm{y})\equiv d\pmod{2}$, we conclude that $a=\bar{b}$.

If $\left|I_2(\bm{x})\cap I_2(\bm{y})\right|=n+5$, then
 $\bm{v}\in\{(a\bar{a})^{i}(\bar{a}a)^{j},$ $(a\bar{a})^{i}aa(\bar{a}a)^{j}\}$ for some $i,j\ge 0$ by \Cref{thm_classification1}(\Rnum{1}). Since $\bm{x}\in R(n,3,\frac{P}{3})$, we have $2i,2j\le \frac{P}{3}$.  From \Cref{eq_inv}, we know that $$\left|\Inv{\bm{x}}-\Inv{\bm{y}}\right|=\left|\Inv{a\bar{a}\bm{v}\bar{a}}-\Inv{\bar{a}\bm{v}\bar{a}a}\right|=i+j+2.$$
So $i+j+2\equiv 0\pmod{1+P}$ by the definition of $\mathcal{C}$. Then $1+P\le i+j+2\leq \frac{P}{3}+2$, which implies that $P\le 1$, a contradiction.

If $\left|I_2(\bm{x})\cap I_2(\bm{y})\right|= n+4$, then  $\bm{v}$ is one of the four forms in \Cref{tb_thmC1} under the case $a=\bar{b}$ by \Cref{thm_classification1}(\Rnum{2}). For each case, we can deduce that $P\leq 3$, thus a contradiction. For example, if $\bm{v}=(a\bar{a})^i(\bar{a}a\bar{a})^j(\bar{a}a)^{k}$ for some $i\ge 0,j\ge 1,k\ge 0$, then $|\Inv{\bm{x}}-\Inv{\bm{y}}|=i+2j+k+2 \equiv 0\pmod{1+P}$. On the other hand, since $\bm{x}\in R(n,3,\frac{P}{3})$, we have $2i\leq \frac{P}{3}$, $3j\leq \frac{P}{3}$, and $2k\leq \frac{P}{3}$. So  $1+P\leq i+2j+k+2\leq \frac{5P}{9}+2$, which implies $P\leq 2$, a contradiction.

So we conclude that $\mathcal{C}(n;c,d)$ is an $(n,n+4;B^{I(2)})$-reconstruction code.
Finally, we can choose $P=3\lceil \log_2(n)\rceil+12$. Then $\left|R(n,3,\frac{P}{3})\right|\ge 2^{n-1}$ by \Cref{lem_zq}. Note that for different pairs $(c,d)\in \mathbb{Z}_{1+P}\times \mathbb{Z}_2$, the codes $\mathcal{C}(n;c,d)$ are disjoint from each other. So by pigeonhole principle, there must be some $c$ and $d$ such that $|\mathcal{C}(n;c,d)|\ge\frac{2^{n-1}}{2(1+P)}$, which has redundancy $\log_2(P+1)+2=\log_2\log_2(n)+O(1)$ for $n$ large enough.
\end{IEEEproof}

Obviously, the code $\mathcal{C}(n;c,d)$ in \Cref{thm_construction1} is also an $(n,n+5;B^{I(2)})$-reconstruction code. Then combining with \Cref{prop_summary1}, we have
$\rho(n,N;B^{I(2)})=\log_2\log_2(n)+\Theta(1)$ for $N\in\{n+4,n+5\}$. However, we can do a little bit better for $N=n+5$. Under the same notations, let   \[\mathcal{C}_2(n;c,d)=\{\bm{x}\in R(n,2,\frac{2P}{3}):  \Inv{\bm{x}}\equiv c\pmod{1+P}, wt_H(\bm{x})\equiv d\pmod{2}\}\}.\] Then we can show that $\mathcal{C}_2(n;c,d)$ is an $(n,n+5;B^{I(2)})$-reconstruction code by the similar argument in \Cref{thm_construction1}.  Furthermore, if we set $2P=3\lceil \log_2(n)\rceil +9$, then $\mathcal{C}_2(n;c,d)$ has redundancy at most $\log_2(P+1)+2$ for some choice of $c$ and $d$. Since this $P$ is about a half of the $P$ in \Cref{thm_construction1}, we know that the redundancy of $\mathcal{C}_2(n;c,d)$ is about  one bit smaller than that of $\mathcal{C}(n;c,d)$.

\section{Reconstruction codes for $N\in\{5,6\}$}\label{sec_N5}
In this section, we provide reconstruction codes for $N\in\{5,6\}$ with redundancy as small as possible. It seems that we can also construct a code by characterizing the structures of two sequences $\bm{x}, \bm{y}$ such that $\left|I_2(\bm{x})\cap I_2(\bm{y})\right|=5$ or $6$, as we did in the last section. However, it will be a long and tedious task. So the construction we use here is quite different from the one for $N=n+4$.

 In \cite[Theorem 2]{wz5}, Sima et al. constructed a class of $2$-deletion correcting codes with redundancy $7\log_2(n)+o(\log_2 (n))$, by generalizing the VT code with  higher order parity checks of some indicator vectors of the original sequences. Inspired by their constructions, we will present an $(n,5;B^{I(2)})$-reconstruction code with redundancy $\log_2(n)+O(\log_2 \log_2(n))$ by utilizing five different reads. This improves the trivial upper bound $7\log_2(n)+o(\log_2 (n))$ to $\log_2(n)+O(\log_2 \log_2(n))$, which is tight in terms of the leading term.
We first recall the construction in \cite[Theorem 2]{wz5},  which needs the following notations.

For  $\bm{x}=x_1x_2\ldots x_n\in\Sigma_2^n$ ( $n\ge 2$) and $a,b\in\Sigma_2$, the $ab$-indicator $\mathbbm{1}_{ab}(\bm{x})\in\Sigma_2^{n-1}$ of $\bm{x}$ is defined as follows:
$$
\mathbbm{1}_{ab}(\bm{x})_i=
\begin{cases}
  1, & \mbox{if } x_i=a\text{ and }x_{i+1}=b, \\
  0, & \mbox{otherwise}.
\end{cases}
$$
For example, if $\bm{x}=10011010$, then $\mathbbm{1}_{10}(\bm{x})=1000101$ and $\mathbbm{1}_{01}(\bm{x})=0010010$. Note that the $10$- and $01$-indicators of any sequence do not contain consecutive ones. Define the following integer vectors of length $n-1$:
\begin{equation}\label{eq:m}
\begin{array}{l}
 \bm{m}^{(0)} \triangleq (1,2,\ldots,n-1),\\
 \bm{m}^{(1)} \triangleq \left(1,1+2,1+2+3,\ldots,\frac{n(n-1)}{2}\right),\\
 \bm{m}^{(2)} \triangleq \left(1^2,1^2+2^2,1^2+2^2+3^{2},\ldots,\frac{(n-1)n(2n-1)}{6}\right).
\end{array}
\end{equation}
In other words, the $i$th components of $\bm{m}^{(0)}$, $\bm{m}^{(1)}$ and $\bm{m}^{(2)}$ are $i$, $i(i+1)/2$ and $i(i+1)(2i+1)/6$, respectively.
Then given $\bm{x}\in\Sigma_2^n$, the higher order parity checks for $\mathbbm{1}_{10}(\bm{x})$ and $\mathbbm{1}_{01}(\bm{x})$ are defined as
\begin{equation}\label{eq:fh}
\begin{array}{l}
  f(\bm{x})=\left(\mathbbm{1}_{10}(\bm{x})\cdot\bm{m}^{(0)}\pmod{2n},~~\mathbbm{1}_{10}(\bm{x})\cdot\bm{m}^{(1)}\pmod{n^2},~~\mathbbm{1}_{10}(\bm{x})\cdot\bm{m}^{(2)}\pmod{n^3}\right),\\
  h(\bm{x})=\left(wt_H\left(\mathbbm{1}_{01}(\bm{x})\right)\pmod{3},~~\mathbbm{1}_{01}(\bm{x})\cdot\bm{m}^{(1)}\pmod{2n}\right),
\end{array}
\end{equation}
where $\cdot$ denotes the inner product over the integers.
By applying the above parity checks, Sima et al. proved the following result.
\begin{thm}\cite[Theorem 2]{wz5}\label{lem_2deletioncode}
For any $a_1\in\mathbb{Z}_{2n}$, $a_2\in\mathbb{Z}_{n^2}$, $a_3\in\mathbb{Z}_{n^3}$, $a_4\in\mathbb{Z}_3$ and $a_5\in\mathbb{Z}_{2n}$, let \[\mathcal{D}(a_1,a_2,a_3,a_4,a_5)=\left\{\bm{x}\in\Sigma_2^n:f(\bm{x})=
\left(a_1,a_2,a_3\right),h(\bm{x})=\left(a_4,a_5\right)\right\}.\]
Then $\mathcal{D}(a_1,a_2,a_3,a_4,a_5)$ is a $2$-deletion (or insertion) correcting code.
\end{thm}
Clearly, the redundancy of $\mathcal{D}(a_1,a_2,a_3,a_4,a_5)$ is at most $7\log_2(n)+O(1)$ for some choice of $a_1$, $a_2$, $a_3$, $a_4$ and $a_5$. Although this implies an $(n,5;B^{I(2)})$-reconstruction code, the redundancy is large. To reduce the redundancy, we note that the parity checks in the construction of  \Cref{lem_2deletioncode} are applied to the whole sequence of length $n$ to correct two random insertion errors. However, as will be shown later, the two random insertion errors are located in a short subword (say of length $L$) of the original sequence, if we impose some constraint (belonging to $R(n,3,P)$) on the sequences. So we can apply the parity checks in \Cref{eq:fh} only on this subword of length $L$ to combat the two insertion errors, which will greatly reduce the redundancy of the code if $L=o(n)$.

\vspace{0.5cm}

Now assume that $\bm{x} \neq\bm{y}\in\Sigma_2^{n}$ ($n\ge 4$) and $|I_2(\bm{x})\cap I_2(\bm{y})|\leq 6$, then
 $|I_1(\bm{x})\cap I_1(\bm{y})|=0$ by \Cref{lem_max2}, and hence $d_H(\bm{x},\bm{y})\ge 2$ by \Cref{lem_typA}. So we can always assume that $$
  \bm{x}=\bm{u}a\bm{d}b\bm{w} \text{ and }
  \bm{y}=\bm{u}\bar{a}\bm{e}\bar{b}\bm{w}
$$ for some $\bm{u},\bm{w},\bm{d},\bm{e}\in\Sigma_2^{*}$ and $a,b\in\Sigma_2$. Further, $a\bm{d}\ne\bm{e}\bar{b}$ and $\bm{d}b\ne\bar{a}\bm{e}$, since otherwise $|I_1(\bm{x})\cap I_1(\bm{y})|\neq0$. As in \Cref{lem_reduced}, we have
\begin{equation}\label{eq:I2}
  I_2(\bm{x})\cap I_2(\bm{y})=\bm{u}\left(I_2(a\bm{d}b)\cap I_2(\bar{a}\bm{e}\bar{b})\right)\bm{w}.
\end{equation}
In particular, $\left|I_2(\bm{x})\cap I_2(\bm{y})\right|=\left|I_2(a\bm{d}b)\cap I_2(\bar{a}\bm{e}\bar{b})\right|$. The proof of \Cref{eq:I2} is similar to that of \Cref{lem_reduced} and thus omitted.

\begin{figure}[!t]
\centering
{
\subfigure[inserted symbols are in the same segment]
{
\begin{minipage}[H]{0.45\textwidth}
\centering
\begin{tikzpicture}
\draw[ultra thick] (0,0)--(6,0);
\draw (0,0)--(0,0.1);
\draw (1,0)--(1,0.1);
\draw (2.5,0)--(2.5,0.1);
\draw (3.5,0)--(3.5,0.1);
\draw (5,0)--(5,0.1);
\draw (6,0)--(6,0.1);
\draw[loosely dotted,thick] (1.1,0.15)--(2.4,0.15);
\draw[loosely dotted,thick] (3.6,0.15)--(4.9,0.15);
\draw[decorate,decoration=brace] (0,0.15)--(1,0.15)node[above,midway]{$m$};
\draw[decorate,decoration=brace] (2.5,0.15)--(3.5,0.15)node[above,midway]{$m$};
\draw[decorate,decoration=brace] (5,0.15)--(6,0.15)node[above,midway]{$\le m$};
\draw (2.7,0)node[red,scale=1.2]{$\star$};
\draw (3.3,0)node[red,scale=1.2]{$\star$};
\end{tikzpicture}
\end{minipage}
}
\subfigure[inserted symbols are in two adjacent segments]
{
\begin{minipage}[H]{0.45\textwidth}
\centering
\begin{tikzpicture}
\draw[ultra thick] (0,0)--(7,0);
\draw (0,0)--(0,0.1);
\draw (1,0)--(1,0.1);
\draw (2.5,0)--(2.5,0.1);
\draw (3.5,0)--(3.5,0.1);
\draw (4.5,0)--(4.5,0.1);
\draw (6,0)--(6,0.1);
\draw (7,0)--(7,0.1);
\draw[loosely dotted,thick] (1.1,0.15)--(2.4,0.15);
\draw[loosely dotted,thick] (4.6,0.15)--(5.9,0.15);
\draw[decorate,decoration=brace] (0,0.15)--(1,0.15)node[above,midway]{$m$};
\draw[decorate,decoration=brace] (2.5,0.15)--(3.5,0.15)node[above,midway]{$m$};
\draw[decorate,decoration=brace] (3.5,0.15)--(4.5,0.15)node[above,midway]{$m$};
\draw[decorate,decoration=brace] (6,0.15)--(7,0.15)node[above,midway]{$\le m$};
\draw (3,0)node[red,scale=1.2]{$\star$};
\draw (4,0)node[red,scale=1.2]{$\star$};
\end{tikzpicture}
\end{minipage}
}
\caption{Stars represent the inserted symbols}
\label{fig_segments}
}
\end{figure}
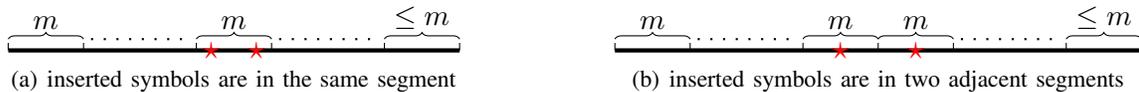

By \Cref{eq:I2}, we know that if  $|I_2(\bm{x})\cap I_2(\bm{y})|\in \{5,6\}$, then any sequence in $I_2(\bm{x})\cap I_2(\bm{y})$ can be obtained from $\bm{x}$ by inserting two symbols in $a\bm{d}b$, or from $\bm{y}$ by inserting two symbols in $\bar{a}\bm{e}\bar{b}$. Therefore, the idea is: if $\left|a\bm{d}b\right|=\left|\bar{a}\bm{e}\bar{b}\right|\le m$ for some $m=o(n)$, then by partitioning $\bm{x}$ (or $\bm{y}$) into non-overlapping segments of length $m$ (see \Cref{fig_segments}, the rightmost segment has length at most $m$), we can see that
the two inserted symbols are located in at most two adjacent segments. Thus we can apply the higher order parity checks in \Cref{eq:fh} to the subword of length $L=2m$. Intuitively, we can deem that the redundancy can be greatly reduced. 

To bound the length of $a\bm{d}b$ or $\bar{a}\bm{e}\bar{b}$, we restrict our sequences to be within $R(n,\ell,P)$, which is of size at least $2^{n-1}$ when $P= \lceil \log_2(n)\rceil +\ell+1$ by \Cref{lem_zq}.
Suppose $\bm{x},\bm{y}\in R(n,\ell,P)$. If we can prove that $a\bm{d}b$ and $\bar{a}\bm{e}\bar{b}$ are both concatenation of a finite number of sequences of period at most $\ell$, then $\left|a\bm{d}b\right|$ and $\left|\bar{a}\bm{e}\bar{b}\right|$ can be upper bounded by  $m=O(P)=o(n)$. Next, we will show that this is true.

Let $S= I_2(a\bm{d}b)\cap I_2(\bar{a}\bm{e}\bar{b})$, where $a\bm{d}\ne\bm{e}\bar{b}$ and $\bm{d}b\ne\bar{a}\bm{e}$. Then $S=S^a_b\cup S^a_{\bar{b}}\cup S_b^{\bar{a}}\cup S^{\bar{a}}_{\bar{b}}$, where
\begin{equation*}
  \begin{array}{l}
    S^a_b=a\left(I_2(\bm{d})\cap\{\bar{a}\bm{e}\bar{b}\}\right)b,\\
    S^a_{\bar{b}}=a\left(I_1(\bm{d}b)\cap I_1(\bar{a}\bm{e})\right)\bar{b},\\
    S^{\bar{a}}_{b}=\bar{a}\left(I_1(a\bm{d})\cap I_1(\bm{e}\bar{b})\right)b,\\
    S^{\bar{a}}_{\bar{b}}=\bar{a}\left(\{a\bm{d}b\}\cap I_2(\bm{e})\right)\bar{b}.
  \end{array}
\end{equation*}
It is clear that $S^a_b$, $S^a_{\bar{b}}$ $S_b^{\bar{a}}$ and $S^{\bar{a}}_{\bar{b}}$ are mutually disjoint, and each has size at most two. Suppose that $|S|=\left|I_2(\bm{x})\cap I_2(\bm{y})\right|\in \{5,6\}$. Then we must have:
\begin{equation}\label{eq_S}
  \left|S_{\bar{b}}^{a}\right|=2 \text{ and } \left|S_{b}^{\bar{a}}\right|\ge 1;\text{ or }
  \left|S_{b}^{\bar{a}}\right|=2 \text{ and } \left|S_{\bar{b}}^{a}\right|\ge 1.
\end{equation}

For the former case of \Cref{eq_S}, $\bm{d}b$ and $\bar{a}\bm{e}$ are Type-A confusable, i.e., there exist some $\bm{\alpha},\bm{\beta}\in\Sigma_2^{*}$ such that
\begin{equation}\label{eq_parttypA}
\bm{d}b=\bm{\alpha}\bm{w}\bm{\beta} \text{ and } \bar{a}\bm{e}=\bm{\alpha}\overline{\bm{w}}\bm{\beta},
\end{equation}
where $\bm{w}\in\Sigma_2^{*}$ is an alternating sequence of length at least one. Since $\left|S_{b}^{\bar{a}}\right|\ge 1$, we conclude that $a\bm{d}$ and $\bm{e}\bar{b}$ are Type-A confusable or Type-B confusable.
With these notations and observations, we characterize the structures of $\bm{\alpha}$ and $\bm{\beta}$ in the following two lemmas for different cases. The proofs are in Appendix~\ref{appendix5}.
\begin{lem}\label{lem_part1}
  Suppose that $\bm{d}b$ and $\bar{a}\bm{e}$ are Type-A confusable, and that $a\bm{d}$ and $\bm{e}\bar{b}$ are Type-A confusable. Then $\bm{\alpha}=\bm{p}_1^{\prime}\bm{p}_2^{\prime}$ and $\bm{\beta}=\bm{p}_1\bm{p}_2$ for some $\bm{p}_1,\bm{p}_2,\bm{p}_1^{\prime},\bm{p}_2^{\prime}\in\Sigma_2^{*}$, where the four sequences $\bm{p}_1,\bm{p}_2,\bm{p}_1^{\prime}$ and $\bm{p}_2^{\prime}$ all have period at most $2$.
\end{lem}

\begin{lem}\label{lem_part2}
  Suppose that $\bm{d}b$ and $\bar{a}\bm{e}$ are Type-A confusable, and that $a\bm{d}$ and $\bm{e}\bar{b}$ are Type-B confusable. Then $\bm{\alpha}=\bm{p}_1^{\prime}\bm{p}_2^{\prime}\bm{p}_3^{\prime}$ and $\bm{\beta}=\bm{p}_1\bm{p}_2\bm{p}_3$ for some $\bm{p}_1,\bm{p}_2,\bm{p}_3,\bm{p}_1^{\prime},\bm{p}_2^{\prime},\bm{p}_3^{\prime}\in\Sigma_2^{*}$, where the six sequences $\bm{p}_1,\bm{p}_2,\bm{p}_3,\bm{p}_1^{\prime},\bm{p}_2^{\prime}$ and $\bm{p}_3^{\prime}$ all have period at most $3$.
\end{lem}

For the latter case of \Cref{eq_S}, we have $a\bm{d}$ and $\bm{e}\bar{b}$ are Type-A confusable by $\left|S_{b}^{\bar{a}}\right|=2$, and $\bm{d}b$ and $\bar{a}\bm{e}$ are Type-A (or B) confusable by $\left|S_{\bar{b}}^{a}\right|\ge 1$. As in \Cref{eq_parttypA}, we can let $a\bm{d}=\bm{\alpha}\bm{w}\bm{\beta}$ and  $\bm{e}\bar{b}=\bm{\alpha}\overline{\bm{w}}\bm{\beta}$, and show that \Cref{lem_part1} and \Cref{lem_part2} are both true for this case.

Now suppose that $\bm{x},\bm{y}\in R(n,3,P)$, and $\left|I_2(\bm{x})\cap I_2(\bm{y})\right|\in \{5,6\}$. Then \Cref{lem_part1} and \Cref{lem_part2} imply that
\begin{equation}\label{eq:7p}
  \left|a\bm{d}b\right|=\left|\bar{a}\bm{e}\bar{b}\right|\le 7P+1.
\end{equation} So in the rest of this subsection, we let $m\triangleq 7P+1$. For convenience, we always assume that $m\mid n$, that is, each segment in \Cref{fig_segments} has length $m$. All results in this subsection can be extended to the case $m\nmid n$ by a truncated method, see \Cref{rmk_extension} later. Before giving the construction of an  $(n,5;B^{I(2)})$-reconstruction code, we define the following notations.

For $\bm{x}\in\Sigma_2^{n}$ with $n=sm$, partition $\bm{x}$ into $s$ segments of length $m$ as in \Cref{fig_segments}. For any two consecutive segments, we define the parity checks on this subword of length $2m$ as in \Cref{eq:fh} as follows. For each $k\in \{0,1,\ldots,s-2\}$, let $\mathbbm{1}^{(k)}_{10}(\bm{x})\triangleq\mathbbm{1}_{10}(\bm{x}[km+1:km+2m])$ and $\mathbbm{1}^{(k)}_{01}(\bm{x})\triangleq\mathbbm{1}_{01}(\bm{x}[km+1:km+2m])$, which are the indicators of the $k$-th segment of $\bm{x}$.
Define
\begin{equation*}
  \begin{array}{l}
  f_{k}(\bm{x})=\left(\mathbbm{1}^{(k)}_{10}(\bm{x})\cdot\bm{m}^{(0)}\pmod{4m},~~
 \mathbbm{1}^{(k)}_{10}(\bm{x})\cdot\bm{m}^{(1)}\pmod{4m^2},~~
  \mathbbm{1}^{(k)}_{10}(\bm{x})\cdot\bm{m}^{(2)}\pmod{8m^3}\right),\\
  h_{k}(\bm{x})=\left(wt_H\left(\mathbbm{1}^{(k)}_{01}(\bm{x})\right)\pmod{3},~~\mathbbm{1}^{(k)}_{01}(\bm{x})\cdot\bm{m}^{(0)}\pmod{4m}\right),
\end{array}
\end{equation*}
where $\bm{m}^{(i)}$ is defined in \Cref{eq:m} but with length $2m-1$.
If we let $\bm{z}=\bm{x}[km+1:km+2m]$, then it is easy to see that $f_k(\bm{x})=f(\bm{z})$ and $h_k(\bm{x})=h(\bm{z})$. Next, we will sum up these $f_k$'s and $h_k$'s according to the parity of $k$. Let $K_e$ be the set of even integers and let $K_o$ be the set of odd integers in  $\{0,1,\ldots,s-2\}$, respectively. It is clear that if $k_1,k_2\in K_e$ (or $K_o$) and $k_1\ne k_2$, then the two intervals $[k_1m+1,k_1m+2m]$ and $[k_2m+1,k_2m+2m]$ are disjoint.
Define \[\tilde{f}_{e}(\bm{x})\triangleq\sum_{k\in K_{e}}f_{k}(\bm{x}) \text{ and } \tilde{f}_{o}(\bm{x})\triangleq\sum_{k\in K_{o}}f_{k}(\bm{x})\] with the summation over $\mathbb{Z}_{4m}\times\mathbb{Z}_{4m^2}\times\mathbb{Z}_{8m^3}$, and define \[\tilde{h}_{e}(\bm{x})\triangleq \sum_{k\in K_{e}}h_{k}(\bm{x}) \text{ and } \tilde{h}_{o}(\bm{x})\triangleq \sum_{k\in K_{o}}h_{k}(\bm{x})\] with the summation over $\mathbb{Z}_3\times\mathbb{Z}_{4m}$. Now we are ready to give our construction.

\begin{thm}\label{thm_construction4}\textup{($N=5$)}
Let $n$ and $P$ be integers such that $m=7P+1<n$ and $m\mid n$. For any $\bm{a}$ $=$ $(a_1,a_2,a_3,a_4,a_5)$, $\bm{b}=(b_1,b_2,b_3,b_4,b_5)$ $\in$ $\mathbb{Z}_{4m}\times\mathbb{Z}_{4m^2}\times\mathbb{Z}_{8m^3}\times\mathbb{Z}_3\times\mathbb{Z}_{4m}$, and $a\in\mathbb{Z}_{n+1}$, let $\mathcal{E}(n;a,\bm{a},\bm{b})$ be the set of all $\bm{x}=x_1\cdots x_n\in\Sigma_2^n$ such that the following four conditions hold:
  \begin{itemize}
    \item $\mathop{\sum}\limits_{i=1}^{n}ix_i=a\pmod{n+1}$;
    \item $\left(\tilde{f}_{e}(\bm{x}),\tilde{h}_{e}(\bm{x})\right)=\bm{a}$;
    \item $\left(\tilde{f}_{o}(\bm{x}),\tilde{h}_{o}(\bm{x})\right)=\bm{b}$; and
    \item $\bm{x}\in R(n,3,P)$.
  \end{itemize}
Then $\mathcal{E}(n;a,\bm{a},\bm{b})$ is an $(n,5;B^{I(2)})$-reconstruction code. Furthermore, if $P=\lceil \log_2(n)\rceil +4$, then
  $\mathcal{E}(n;a,\bm{a},\bm{b})$ has redundancy at most $19+2\log_2(3)+\log_2(n+1)+14\log_2(7P+1)=\log_2(n)+14\log_2\log_2(n)+O(1)$ for some choice of $\bm{a}$, $\bm{b}$, and $a$.
\end{thm}
\begin{IEEEproof}
According to \Cref{eq_VTcode}, the first condition implies that $\mathcal{E}(n;a,\bm{a},\bm{b})$ is a subset of a VT code. So $I_1(\bm{x})\cap I_1(\bm{y})=\emptyset$,  and hence $d_H(\bm{x},\bm{y})\ge 2$ for any distinct $\bm{x},\bm{y}$.  Assume that $$
  \bm{x}=\bm{u}a\bm{d}b\bm{w} \text{ and }
  \bm{y}=\bm{u}\bar{a}\bm{e}\bar{b}\bm{w}
$$ for some $\bm{u},\bm{w},\bm{d},\bm{e}\in\Sigma_2^{*}$ and $a,b\in\Sigma_2$.
By \Cref{lem_max2}, $|I_2(\bm{x})\cap I_2(\bm{y})|\leq 6$. We will show that $\left|I_2(\bm{x})\cap I_2(\bm{y})\right|<5$ by contradiction.

Assume that $\left|I_2(\bm{x})\cap I_2(\bm{y})\right|\ge 5$. Then by \Cref{eq:7p} we can partition the sequences $\bm{x}$ and $\bm{y}$  into $s=n/m$ segments each of length $m$, such that the two inserted symbols in the common supersequence are located in at most two adjacent segments containing $a\bm{d}b$ or $\bar{a}\bm{e}\bar{b}$. See \Cref{fig_segments} for illustrations.
This means that there exists some $k\in \{0,1,\ldots,s-2\}$ such that  $a\bm{d}b$ and $\bar{a}\bm{e}\bar{b}$ are subwords of $\bm{x}[km+1:km+2m]$ and $\bm{y}[km+1:km+2m]$, respectively. Without loss of generality, we assume $k\in K_e$ and the proof is the same for the case $k\in K_o$.
Let $\bm{x}^{\prime}\triangleq\bm{x}[km+1:km+2m]$ and $\bm{y}^{\prime}\triangleq\bm{y}[km+1:km+2m]$.  Then $f_{k}(\bm{x})=f(\bm{x}^{\prime})$, $f_{k}(\bm{y})=f(\bm{y}^{\prime})$, $h_{k}(\bm{x})=h(\bm{x}^{\prime})$ and $h_{k}(\bm{y})=h(\bm{y}^{\prime})$.
 By the choice of $k$, we know that $\bm{x}[k^{\prime}m+1:(k^{\prime}+2)m]$ $=$ $\bm{y}[k^{\prime}m+1:(k^{\prime}+2)m]$ for each $k^{\prime}\in K_e\setminus\{k\}$, and hence $f_{k^{\prime}}(\bm{x})=f_{k^{\prime}}(\bm{y})$ and $h_{k^{\prime}}(\bm{x})=h_{k^{\prime}}(\bm{y})$.
Therefore, we have
\begin{equation*}
  \begin{array}{l}
    \tilde{f}_{e}(\bm{x})-\tilde{f}_{e}(\bm{y})=f_{k}(\bm{x})-f_{k}(\bm{y})=f(\bm{x}^{\prime})-f(\bm{y}^{\prime}),\\
    \tilde{h}_{e}(\bm{x})-\tilde{h}_{e}(\bm{y})=h_{k}(\bm{x})-h_{k}(\bm{y})=h(\bm{x}^{\prime})-h(\bm{y}^{\prime}).
  \end{array}
\end{equation*}
 The second condition in this theorem implies $\tilde{f}_{e}(\bm{x})=\tilde{f}_{e}(\bm{y})$ and $\tilde{h}_{e}(\bm{x})=\tilde{h}_{e}(\bm{y})$, i.e., $f(\bm{x}^{\prime})=f(\bm{y}^{\prime})$ and $h(\bm{x}^{\prime})=h(\bm{y}^{\prime})$. By \Cref{lem_2deletioncode}, we know that $\left|I_2(\bm{x}^{\prime})\cap I_2(\bm{y}^{\prime})\right|=0$. However, by \Cref{eq:I2}, $\left|I_2(\bm{x}^{\prime})\cap I_2(\bm{y}^{\prime})\right|$ $=$ $\left|I_2(\bm{x})\cap I_2(\bm{y})\right|$ $\geq 5$, a contradiction. Thus the code $\mathcal{E}(n;a,\bm{a},\bm{b})$ is indeed an $(n,5;B^{I(2)})$-reconstruction code.

Finally, we choose $P=\lceil \log_2(n)\rceil +4$, which implies that $\left|R(n,3,P)\right|\ge 2^{n-1}$ by \Cref{lem_zq}. So there must be some $a\in\mathbb{Z}_{n+1}$, $\bm{a}$, $\bm{b}$ $\in$ $\mathbb{Z}_{4m}\times\mathbb{Z}_{4m^2}\times\mathbb{Z}_{8m^3}\times\mathbb{Z}_3\times\mathbb{Z}_{4m}$, such that $|\mathcal{E}(n;a,\bm{a},\bm{b})|\ge\frac{2^{n-1}}{(n+1)M^2}$,  where $M={4m}\cdot{4m^2}\cdot{8m^3}\cdot3\cdot{4m}=3\cdot 2^9\cdot m^7$. By computing the redundancy, we complete the proof.
\end{IEEEproof}

\begin{rmk}
If we replace the fourth condition in \Cref{thm_construction4} with $\bm{x}\in R(n,2,P)$, and take $P=\left\lceil\log_2(n)\right\rceil+3$ and  $m=5P+1$, then we will get an $(n,6;B^{I(2)})$-reconstruction code with redundancy at most $\log_2(n)+14\log_2\log_2(n)+O(1)$, the constant term of which is slightly smaller than that of the redundancy of the code given in \Cref{thm_construction4}.
\end{rmk}
\begin{rmk}\label{rmk_extension}
  Now we explain how to extend the construction in Theorem~\ref{thm_construction4} to the case $m\nmid n$. Let $P=\lceil \log_2(n)\rceil +4$ and $m=7P+1$ as before. Suppose that $\bar{n}$ be the smallest integer such that $\bar{n}>n$ and $m\mid \bar{n}$. For each $\bm{x}\in\Sigma_2^n$, let $\bar{\bm{x}}$ be the word of length $\bar{n}$ by appending $\bar{n}-n$ zeros at the end of $\bm{x}$. Then in \Cref{thm_construction4}, modify the second and third conditions by replacing $\bm{x}$ with $\bar{\bm{x}}$, and keep the other two conditions unchanged. By almost the same arguments, we can show that there exists an $(n,5;B^{I(2)})$-reconstruction code with redundancy  at most $\log_2(n)+14\log_2\log_2(n)+O(1)$.
\end{rmk}

When $N\in\{5,6\}$, an $(n,N;B^{I(2)})$-reconstruction code must be an $(n,1;B^{I(1)})$-reconstruction code. So the lower bound of $\rho(n,N;B^{I(2)})$ is $\log_2(n)+\Omega(1)$ (see \Cref{eq_VTcode}). Therefore, the redundancy of the code in
\Cref{thm_construction4} has optimal leading term.
  It is worth noting that the idea in this section can be applied to all $N\le 6$. The main task is to upper bound the length of $a\bm{d}b$ and $\bar{a}\bm{e}\bar{b}$ under some restrictions. We only accomplished this task when $N\in\{5,6\}$. When $N\le 4$, the analysis will be more complicated.

To conclude this section, we summarize the main results into the following theorem.
\begin{thm}
\begin{equation*}
\rho(n,N;B^{I(2)})=
\begin{cases}
  \Theta(\log_2(n)), & \mbox{if } 1\le N\le 4, \\
   \log_2(n)+O(\log_2\log_2(n)), & \mbox{if } N=5, 6,\\
  \log_2(n)+\Theta(1), & \mbox{if } 6<N\le n+3, \\
  \log_2\log_2(n)+\Theta(1), & \mbox{if } n+4\leq N\le 2n+4, \\
 0, & \mbox{if } N>2n+4. \\
\end{cases}
\end{equation*}
\end{thm}

\section{Conclusion}\label{sec_conclusion}

In this paper, we  study the redundancy of reconstruction codes for channels with exactly two insertions. Let $N$ be the number of given channels and let $n$ be the sequence length. It turns out that the nontrivial cases are $1\le N\le 6$ and $N\in\{n+4,n+5\}$. For $N=n+4$ and $N=5$, our constructions of codes are both explicit, and the ideas are quite different.  When $N=n+4$, the construction is based on characterizations of two sequences when the intersection size of their $2$-insertion balls is exactly $n+4$ or $n+5$. When $N=5$, the construction is based on higher order parity checks on short subwords. Consequently, for all $N\ge 5$, we construct codes which are asymptotically optimal in terms of redundancy.

For general  $t$-insertions (with $t> 2$ being a constant), we have the following result as a generalization of \Cref{lem_basecase2}.
\begin{lem}\label{thm_tinsertion}
 Let $n\ge 3$ and $\bm{x},\bm{y}\in\Sigma_2^n$. Suppose that  $\left|I_{1}(\bm{x})\cap I_{1}(\bm{y})\right|=1$, then
  \begin{align*}
   \left|I_{t}(\bm{x})\cap I_{t}(\bm{y})\right|\le I_2(n+1,t-1)+N_2^{+}(n-1,t-1)
  \end{align*}
  for any $t\ge 2$.
\end{lem}

The proof of \Cref{thm_tinsertion} is in Appendix~\ref{appendix6}. By \Cref{Nl=0} and \Cref{Nl=1}, we can see that $I_2(n+1,t-1)+N_2^{+}(n-1,t-1)=\frac{n^{t-1}}{(t-1)!}+O(n^{t-2})$.
Thus \Cref{thm_tinsertion} gives a similar result as in \cite{wz28}: the reconstruction code with two reads for a single insertion \cite{wz2} is able to reconstruct codewords from $N=n^{t-1}/(t-1)!+O(n^{t-2})$ distinct noisy reads. 

For future research, we raise the following questions.
\begin{enumerate}[(1)]
  \item In \Cref{thm_construction4}, we construct an $(n,5;B^{I(2)})$-reconstruction code with redundancy $\log_2(n)+14\log_2\log_2(n)+O(1)$. The leading term $\log_2(n)$ is optimal. But we do not know whether the coefficient of the secondary term is optimal or not. This is left for future study.
  \item Using the notations of \Cref{sec_N5}, we can see that $\left|I_2(\bm{x})\cap I_2(\bm{y})\right|=6$ if and only if $\left|S^a_{\bar{b}}\right|=\left|S^b_{\bar{a}}\right|=2$ and $\left|S^a_{b}\right|=\left|S^{\bar{b}}_{\bar{a}}\right|=1$. In other words, $\left|I_2(\bm{x})\cap I_2(\bm{y})\right|=6$ if and only if
     \begin{equation*}
        \begin{array}{l}
          \bm{d}b\text{ and }\bar{a}\bm{e}\text{ are Type-A confusable},\\
           a\bm{d}\text{ and }\bm{e}\bar{b}\text{ are Type-A confusable},\\
          \bar{a}\bm{e}\bar{b}\in I_2(\bm{d}), \text{ and }
           a\bm{d}b\in I_2(\bm{e}).
        \end{array}
     \end{equation*}
     By these equivalent conditions, we can  determine the explicit structures of $\bm{d}$ and $\bm{e}$. Then similar to \Cref{thm_construction1}, we can construct an $(n,6;B^{I(2)})$-reconstruction code with redundancy $\log_2(n)+2\log_2\log_2(n)+O(1)$. But the proofs are long and tedious. The details of this construction can be provided upon requirement. Again, we do not know whether the secondary term is optimal or not.
  \item Find a way to upper bound the length of $a\bm{d}b$ and $\bar{a}\bm{e}\bar{b}$  in \Cref{sec_N5}. This will be helpful for us to construct $(n,N;B^{I(2)})$-reconstruction codes for $2\le N\le 4$.
\end{enumerate}

\appendices
\section{The proof of \Cref{lem_reduced}}\label{appendix1}
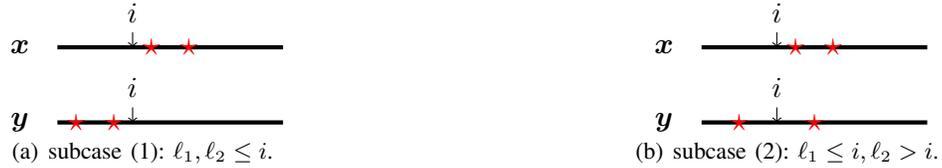
\begin{figure}[!t]
\centering
{
\subfigure[subcase (1): $\ell_1,\ell_2\le i$.]
{
\begin{minipage}[H]{0.45\textwidth}
\centering
\begin{tikzpicture}
\draw[ultra thick] (0,0)--(3,0);
\draw[->] (1,0.2)--(1,0);
\draw (1,0.2) node[above]{$i$};
\draw (1.25,0)node[red,scale=1.2]{$\star$};
\draw (1.75,0)node[red,scale=1.2]{$\star$};
\draw (-0.5,0)node{$\bm{x}$};
\draw[ultra thick] (0,-1)--(3,-1);
\draw[->] (1,-0.8)--(1,-1);
\draw (1,-0.8) node[above]{$i$};
\draw (0.25,-1)node[red,scale=1.2]{$\star$};
\draw (0.75,-1)node[red,scale=1.2]{$\star$};
\draw (-0.5,-1)node{$\bm{y}$};
\end{tikzpicture}
\end{minipage}
}
\subfigure[subcase (2): $\ell_1\le i,\ell_2>i$.]
{
\begin{minipage}[H]{0.45\textwidth}
\centering
\begin{tikzpicture}
\draw[ultra thick] (0,0)--(3,0);
\draw[->] (1,0.2)--(1,0);
\draw (1,0.2) node[above]{$i$};
\draw (1.25,0)node[red,scale=1.2]{$\star$};
\draw (1.75,0)node[red,scale=1.2]{$\star$};
\draw (-0.5,0)node{$\bm{x}$};
\draw[ultra thick] (0,-1)--(3,-1);
\draw[->] (1,-0.8)--(1,-1);
\draw (1,-0.8) node[above]{$i$};
\draw (0.5,-1)node[red,scale=1.2]{$\star$};
\draw (1.5,-1)node[red,scale=1.2]{$\star$};
\draw (-0.5,-1)node{$\bm{y}$};
\end{tikzpicture}
\end{minipage}
}
\caption{Case (3): $k_1,k_2>i$. Stars are the inserted positions.}
\label{fig_case3}
}
\end{figure}

\begin{IEEEproof}
Recall that $\bm{x}=\bm{u}a\bar{a}\bm{v}b\bm{w}$, $\bm{y}=\bm{u}\bar{a}\bm{v}b\bar{b}\bm{w}$ and $\bm{z}=\bm{u}a\bar{a}\bm{v}b\bar{b}\bm{w}$.
For simplicity, we denote $\bm{\widetilde{x}}=a\bar{a}\bm{v}b$, $\bm{\widetilde{y}}=\bar{a}\bm{v}b\bar{b}$, $\bm{\widetilde{z}}=a\bar{a}\bm{v}b\bar{b}$ and $T=I_2(\widetilde{\bm{x}})\cap I_2(\widetilde{\bm{y}})\setminus I_1(\bm{\widetilde{z}})$. Clearly, $\bm{u}I_1(\bm{\widetilde{z}})\bm{w}\subset I_1(\bm{z})$, and the two sets $I_1(\bm{z}), \bm{u}T\bm{w}$ are disjoint. Next we will prove $I_2(\bm{x})\cap I_2(\bm{y})=I_1(\bm{z})\cup \bm{u}T\bm{w}$. It is easy to see that
$$
I_1(\bm{z})\cup \bm{u}T\bm{w}\subseteq I_2(\bm{x})\cap I_2(\bm{y}).
$$
So we only need to prove that the opposite inclusion is true. Since $\bm{u}I_1(\bm{\widetilde{z}})\bm{w}\subset I_1(\bm{z})$, it is sufficient to show $I_2(\bm{x})\cap I_2(\bm{y})$ $\subseteq$ $I_1(\bm{z})\cup \bm{u}\left(I_2(\widetilde{\bm{x}})\cap I_2(\widetilde{\bm{y}})\right)\bm{w}$.
Assume that $\bm{z}^{\prime}\in I_2(\bm{x})\cap I_2(\bm{y})$. Suppose that $k_1\le k_2$ are the two insertion positions in $\bm{x}$,\footnote{This means that two symbols are inserted to the left of $x_{k_1}$ and $x_{k_2}$, respectively, to get $\bm{z}^{\prime}$.} and $\ell_1\le \ell_2$ are the two insertion positions in $\bm{y}$. Let $i$ and $j$ be the leftmost and rightmost indices where $\bm{x}$ and $\bm{y}$ differ (see \Cref{fig_differ}). Comparing $k_1,k_2$ with $i$, we can divide the discussions into three cases: (1) $k_1,k_2\le i$; (2) $k_1\le i,k_2> i$; (3) $k_1,k_2>i$.
\par\noindent
\textbf{Case (1)}: $k_1,k_2\le i$ (\Cref{fig_case1}). Since $x_j\neq y_j$, we must have $\ell_2> j$ in this case, and either $\ell_1> j$ or $\ell_1\le j$.
\par
\underline{Subcase (1)}: If $\ell_1> j$, then by matching positions, we have
$$
\bm{z}^{\prime}\in I_2(\bm{u})a\bar{a}\bm{v}b\bm{w}\cap \bm{u}\bar{a}\bm{v}b\bar{b}I_2(\bm{w})= \bm{u}\left(I_2(\bm{\widetilde{x}})\cap I_2(\bm{\widetilde{y}})\right)\bm{w}.
$$

\underline{Subcase (2)}: If $\ell_1\le j$, then
\begin{align*}
  \bm{z}^{\prime}&\in I_2(\bm{u})a\bar{a}\bm{v}b\bm{w}\cap I_1(\bm{u}\bar{a}\bm{v}b)\bar{b}b\bm{w} \\
  &=\left(I_2(\bm{u})a\bar{a}\bm{v}\cap I_1(\bm{u}\bar{a}\bm{v}b)\bar{b}\right) b\bm{w}.
\end{align*}
For $I_2(\bm{u})a\bar{a}\bm{v}\cap I_1(\bm{u}\bar{a}\bm{v}b)\bar{b}$, we have
$$
I_2(\bm{u})a\bar{a}\bm{v}\cap I_1(\bm{u}\bar{a}\bm{v}b)\bar{b}=\left(I_2(\bm{u})a\bar{a}\bm{v}\cap \bm{u}I_1(\bar{a}\bm{v}b)\bar{b}\right)\cup\left(I_2(\bm{u})a\bar{a}\bm{v}\cap I_1(\bm{u})\bar{a}\bm{v}b\bar{b}\right).
$$
If the second bracket $I_2(\bm{u})a\bar{a}\bm{v}\cap I_1(\bm{u})\bar{a}\bm{v}b\bar{b}$ is not empty, then $a\bar{a}\bm{v}=\bm{v}b\bar{b}$. This holds if and only if $\bm{v}=(a\bar{a})^m$ and $a=b$, or $\bm{v}=(a\bar{a})^ma$ and $a=\bar{b}$, for some $m\ge 0$. Both cases imply that $\bm{x}$ and $\bm{y}$ are Type-A confusable by \Cref{prop_summary}(2), which is a contradiction. Therefore, $I_1(\bm{u})\bar{a}a\bar{a}\bm{v}\cap I_1(\bm{u})\bar{a}\bm{v}b\bar{b}$ is empty and so
$$
\bm{z}^{\prime}\in\left(I_2(\bm{u})a\bar{a}\bm{v}\cap \bm{u}I_1(\bar{a}\bm{v}b)\bar{b}\right)b\bm{w}\subseteq \bm{u}I_1(\bar{a}\bm{v}b)\bar{b}b\bm{w}\subseteq \bm{u}I_2(\bm{\widetilde{y}})\bm{w}.
$$
This means that $\bm{z}^{\prime}\in\bm{u}\left(I_2(\bm{\widetilde{x}})\cap I_2(\bm{\widetilde{y}})\right)\bm{w}$, since $\bm{z}^{\prime}$ is also in $I_2(\bm{u}a\bar{a}\bm{v}b\bm{w})$ ($=I_2(\bm{x})$) by assumption.
\par\noindent
\textbf{Case (2)} (\Cref{fig_case2}): $k_1\le i,k_2> i$. Since $x_i\neq y_i$, we can not have exactly one of $\ell_i$ less than $i$. Thus
we have $\ell_1,\ell_2\le i$ or $\ell_1,\ell_2> i$
 in this case.
\par
\underline{Subcase (1)}: If $\ell_1,\ell_2\le i$ then we must have $k_2> j$ since $x_j\ne y_j$. So we have
\begin{align*}
  \bm{z}^{\prime} & \in I_1(\bm{u})a\bar{a}\bm{v}bI_1(\bm{w})\cap I_2(\bm{u})\bar{a}\bm{v}b\bar{b}\bm{w} \\
   &=I_1(\bm{u})a\bar{a}\bm{v}b\bar{b}\bm{w}\cap I_2(\bm{u})\bar{a}\bm{v}b\bar{b}\bm{w}\\
   &\subseteq I_1(\bm{u}a\bar{a}\bm{v}b\bar{b}\bm{w})=I_1(\bm{z}).
\end{align*}
\par
\underline{Subcase (2)}: If $\ell_1,\ell_2> i$, then
\begin{align*}
  \bm{z}^{\prime} & \in I_1(\bm{u})aI_1(\bar{a}\bm{v}b\bm{w})\cap \bm{u}\bar{a}I_2(\bm{v}b\bar{b}\bm{w}) \\
   &=\bm{u}\bar{a}aI_1(\bar{a}\bm{v}b\bm{w})\cap \bm{u}\bar{a}I_2(\bm{v}b\bar{b}\bm{w}) \\
   &=\bm{u}\bar{a}\left(aI_1(\bar{a}\bm{v}b\bm{w})\cap I_2(\bm{v}b\bar{b}\bm{w})\right).
\end{align*}
Note that $I_1(\bar{a}\bm{v}b\bm{w})=$ $I_1(\bar{a}\bm{v}b)\bm{w}\cup \bar{a}\bm{v}bI_1(\bm{w})$. So
\begin{align*}
aI_1(\bar{a}\bm{v}b\bm{w})\cap I_2(\bm{v}b\bar{b}\bm{w})&=\left(aI_1(\bar{a}\bm{v}b)\bm{w}\cap I_2(\bm{v}b\bar{b}\bm{w})\right)\cup\left(a\bar{a}\bm{v}bI_1(\bm{w})\cap I_2(\bm{v}b\bar{b}\bm{w})\right)\\
&=\left(aI_1(\bar{a}\bm{v}b)\bm{w}\cap I_2(\bm{v}b\bar{b})\bm{w}\right)\cup\left(a\bar{a}\bm{v}bI_1(\bm{w})\cap I_1(\bm{v}b\bar{b})I_1(\bm{w})\right).
\end{align*}
The last equality follows from $I_2(\bm{v}b\bar{b}\bm{w})=I_2(\bm{v}b\bar{b})\bm{w}\cup \bm{v}b\bar{b}I_2(\bm{w})\cup I_1(\bm{v}b\bar{b})I_1(\bm{w})$.
If $a\bar{a}\bm{v}bI_1(\bm{w})\cap I_1(\bm{v}b\bar{b})I_1(\bm{w})$ is not empty, then we have $a\bar{a}\bm{v}b\in I_1(\bm{v}b\bar{b})$, which means $a\bar{a}\bm{v}b=\bm{v}b\bar{b}b$ and so $a\bar{a}\bm{v}=\bm{v}b\bar{b}$. This implies that $\bm{x}$ and $\bm{y}$ are Type-A confusable, which is a contradiction. Therefore, $a\bar{a}\bm{v}bI_1(\bm{w})\cap I_1(\bm{v}b\bar{b})I_1(\bm{w})$ is empty and
$$
\bm{z}^{\prime}\in\bm{u}\bar{a}\left(aI_1(\bar{a}\bm{v}b)\bm{w}\cap I_2(\bm{v}b\bar{b})\bm{w}\right)\subseteq \bm{u}\bar{a}I_2(\bm{v}b\bar{b})\bm{w}\subseteq \bm{u}I_2(\bm{\tilde{y}})\bm{w}.
$$
This means that $\bm{z}^{\prime}\in\bm{u}\left(I_2(\bm{\tilde{x}})\cap I_2(\bm{\tilde{y}})\right)\bm{w}$,  since $\bm{z}^{\prime}$ is also in $I_2(\bm{u}a\bar{a}\bm{v}b\bm{w})$ ($=I_2(\bm{x})$) by assumption.
\par\noindent
\textbf{Case (3)} (\Cref{fig_case3}): $k_1,k_2> i$. In this case, we have $\ell_1,\ell_2\le i$ or $\ell_1\le i,\ell_2>i$. By a process similar to that of the previous two cases, we will get $\bm{z}^{\prime}\in\bm{u}\left(I_2(\bm{\tilde{x}})\cap I_2(\bm{\tilde{y}})\right)\bm{w}$ or $\bm{z}^{\prime}\in I_1(\bm{z})$.

Combining cases (1), (2) and (3), we conclude that \small{$I_2(\bm{x})\cap I_2(\bm{y})\subseteq I_1(\bm{z})\cup \bm{u}T\bm{w}$}. Thus the proof is completed.
\end{IEEEproof}
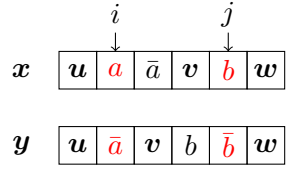
\begin{figure}[!t]
\centering
\begin{tikzpicture}
\draw (0,0) rectangle (0.5,0.5);
\draw (0.5,0) rectangle (1,0.5);
\draw (1,0) rectangle (1.5,0.5);
\draw (1.5,0) rectangle (2,0.5);
\draw (2,0) rectangle (2.5,0.5);
\draw (2.5,0) rectangle (3,0.5);
\path (0.25,0.25)node{$\bm{u}$}
      (0.75,0.25)node{\textcolor{red}{$a$}}
      (1.25,0.25)node{$\bar{a}$}
      (1.75,0.25)node{$\bm{v}$}
      (2.25,0.25)node{\textcolor{red}{$b$}}
      (2.75,0.25)node{$\bm{w}$}
      (-0.5,0.25)node{$\bm{x}$};
\draw (0,-1) rectangle (0.5,-0.5);
\draw (0.5,-1) rectangle (1,-0.5);
\draw (1,-1) rectangle (1.5,-0.5);
\draw (1.5,-1) rectangle (2,-0.5);
\draw (2,-1) rectangle (2.5,-0.5);
\draw (2.5,-1) rectangle (3,-0.5);
\path (0.25,-0.75)node{$\bm{u}$}
      (0.75,-0.75)node{\textcolor{red}{$\bar{a}$}}
      (1.25,-0.75)node{$\bm{v}$}
      (1.75,-0.75)node{$b$}
      (2.25,-0.75)node{\textcolor{red}{$\bar{b}$}}
      (2.75,-0.75)node{$\bm{w}$}
      (-0.5,-0.75)node{$\bm{y}$};
\draw[->](0.75,0.75)--(0.75,0.5);
\draw[->](2.25,0.75)--(2.25,0.5);
\draw (0.75,1)node{$i$};
\draw (2.25,1)node{$j$};
\end{tikzpicture}
\caption{We denote $i$ and $j$ to be the leftmost and rightmost indices where $\bm{x}$ and $\bm{y}$ differ.}
\label{fig_differ}
\end{figure}

\begin{figure}[!t]
\centering
{
\subfigure[subcase (1): $\ell_1,\ell_2>j$.]
{
\begin{minipage}[H]{0.45\textwidth}
\centering
\begin{tikzpicture}
\draw[ultra thick] (0,0)--(3,0);
\draw[->] (1,0.2)--(1,0);
\draw (1,0.2) node[above]{$i$};
\draw (0.25,0)node[red,scale=1.2]{$\star$};
\draw (0.75,0)node[red,scale=1.2]{$\star$};
\draw (-0.5,0)node{$\bm{x}$};
\draw[ultra thick] (0,-1)--(3,-1);
\draw[->] (2,-0.8)--(2,-1);
\draw (2,-0.8) node[above]{$j$};
\draw (2.25,-1)node[red,scale=1.2]{$\star$};
\draw (2.75,-1)node[red,scale=1.2]{$\star$};
\draw (-0.5,-1)node{$\bm{y}$};
\end{tikzpicture}
\end{minipage}
}
\subfigure[subcase (2): $\ell_1\le j,\ell_2>j$.]
{
\begin{minipage}[H]{0.45\textwidth}
\centering
\begin{tikzpicture}
\draw[ultra thick] (0,0)--(3,0);
\draw[->] (1,0.2)--(1,0);
\draw (1,0.2) node[above]{$i$};
\draw (0.25,0)node[red,scale=1.2]{$\star$};
\draw (0.75,0)node[red,scale=1.2]{$\star$};
\draw (-0.5,0)node{$\bm{x}$};
\draw[ultra thick] (0,-1)--(3,-1);
\draw[->] (2,-0.8)--(2,-1);
\draw (2,-0.8) node[above]{$j$};
\draw (2.5,-1)node[red,scale=1.2]{$\star$};
\draw (1.5,-1)node[red,scale=1.2]{$\star$};
\draw (-0.5,-1)node{$\bm{y}$};
\end{tikzpicture}
\end{minipage}
}
\caption{Case (1): $k_1,k_2\le i$. Stars are the inserted positions.}
\label{fig_case1}
}
\end{figure}

\begin{figure}[!t]
\centering
{
\subfigure[subcase (1): $\ell_1,\ell_2\le i$.]
{
\begin{minipage}[H]{0.45\textwidth}
\centering
\begin{tikzpicture}
\draw[ultra thick] (0,0)--(3,0);
\draw[->] (1,0.2)--(1,0);
\draw (1,0.2) node[above]{$i$};
\draw (0.5,0)node[red,scale=1.2]{$\star$};
\draw (1.5,0)node[red,scale=1.2]{$\star$};
\draw (-0.5,0)node{$\bm{x}$};
\draw[ultra thick] (0,-1)--(3,-1);
\draw[->] (1,-0.8)--(1,-1);
\draw (1,-0.8) node[above]{$i$};
\draw (0.25,-1)node[red,scale=1.2]{$\star$};
\draw (0.75,-1)node[red,scale=1.2]{$\star$};
\draw (-0.5,-1)node{$\bm{y}$};
\end{tikzpicture}
\end{minipage}
}
\subfigure[subcase (2): $\ell_1,\ell_2>i$.]
{
\begin{minipage}[H]{0.45\textwidth}
\centering
\begin{tikzpicture}
\draw[ultra thick] (0,0)--(3,0);
\draw[->] (1,0.2)--(1,0);
\draw (1,0.2) node[above]{$i$};
\draw (0.5,0)node[red,scale=1.2]{$\star$};
\draw (1.5,0)node[red,scale=1.2]{$\star$};
\draw (-0.5,0)node{$\bm{x}$};
\draw[ultra thick] (0,-1)--(3,-1);
\draw[->] (1,-0.8)--(1,-1);
\draw (1,-0.8) node[above]{$i$};
\draw (1.25,-1)node[red,scale=1.2]{$\star$};
\draw (1.75,-1)node[red,scale=1.2]{$\star$};
\draw (-0.5,-1)node{$\bm{y}$};
\end{tikzpicture}
\end{minipage}
}
\caption{Case (2): $k_1\le i,k_2>i$. Stars are the inserted positions.}
\label{fig_case2}
}
\end{figure}

\section{Proof of Case (\Rnum{2}) of \Cref{thm_classification1}}\label{appendix2}
\begin{IEEEproof}
(\Rnum{2}). Recall that $\widetilde{\bm{x}}=a\bar{a}\bm{v}$, $\widetilde{\bm{y}}=\bm{v}b\bar{b}$ and $\bm{v}=v_1\cdots v_{l}$, where $l\ge 0$ is the length of $\bm{v}$. In addition, we write $\widetilde{\bm{x}}=\tilde{x}_1\cdots \tilde{x}_{l+2}$ and $\widetilde{\bm{y}}=\tilde{y}_1\cdots \tilde{y}_{l+2}$.

By \Cref{lem_basecase2}, $\left|I_{2}(\bm{x})\cap I_{2}(\bm{y})\right|=n^{\prime}+4$ if and only if $a\bar{a}\bm{v}$ and $\bm{v}b\bar{b}$ are Type-B confusable, but not Type-A confusable.
That is to say,
  \begin{equation}\label{eq_case1}
    \widetilde{\bm{x}}=a\bar{a}\bm{v}=\bm{u}\alpha \bar{\alpha}\widetilde{\bm{v}}\beta\bm{w},\text{  }\widetilde{\bm{y}}=\bm{v}b\bar{b}=\bm{u} \bar{\alpha}\widetilde{\bm{v}}\beta \bar{\beta}\bm{w}
  \end{equation}
or
\begin{equation}\label{eq_case2}
    \widetilde{\bm{x}}=a\bar{a}\bm{v}=\bm{u}\bar{\alpha}\widetilde{\bm{v}}\beta \bar{\beta}\bm{w},\text{  }\widetilde{\bm{y}}=\bm{v}b\bar{b}=\bm{u}\alpha \bar{\alpha}\widetilde{\bm{v}}\beta\bm{w}
  \end{equation}
for some $\bm{u},\widetilde{\bm{v}},\bm{w}\in\Sigma_2^{*}$ and $\alpha,\beta\in\Sigma_2$. Note that $\beta =\bar{\alpha}$ since $a\bar{a}\bm{v}$ and $\bm{v}b\bar{b}$ have the same Hamming weight. Suppose that $|\bm{u}|=s$,$|\bm{w}|=t$, where $s,t\ge 0$. Then $l+2=|a\bar{a}\bm{v}|=s+t+3+|\widetilde{\bm{v}}|$ and so $l\ge s+t+1$.
\par
In order to simplify arguments in the sequel, we let $v_{-1}=a$, $v_0=\bar{a}$, $v_{l+1}=b$ and $v_{l+2}=\bar{b}$. Then it is clear that $\tilde{x}_i=v_{i-2}$ and $\tilde{y}_i=v_i$ for all $1\le i\le l+2$. Both of \Cref{eq_case1} and \Cref{eq_case2} indicate that $\tilde{x}_1\cdots\tilde{x}_s=\tilde{y}_1\cdots\tilde{y}_s$ and $\tilde{x}_{l+3-t}\cdots\tilde{x}_{l+2}=\tilde{y}_{l+3-t}\cdots\tilde{y}_{l+2}$. Therefore, we have the following two identities:
\begin{equation}\label{eq_identity}
  \begin{array}{c}
    v_{-1}\cdots v_{s-2}=v_1\cdots v_s, \\
    v_{l-t+1}\cdots v_{l}=v_{l-t+3}\cdots v_{l+2},
  \end{array}
\end{equation}
for any $s,t\ge 0$ and $l\ge s+t+1$. From the first row in \Cref{eq_identity}, we know that $v_{-1}v_0\cdots v_s$ is the same as that in \Cref{eq_alterv}. And the second row in \Cref{eq_identity} implies that
\begin{equation}\label{eq_vt}
  v_{l-t+1}\cdots v_{l+2}=
  \begin{cases}
    (v_{l-t+1}v_{l-t+2})^{\frac{t+2}{2}}, & \mbox{if } t \text{ is even}, \\
    (v_{l-t+1}v_{l-t+2})^{\frac{t+1}{2}}v_{l-t+1}, &  \mbox{if } t \text{ is odd}.
  \end{cases}
\end{equation}
From \Cref{eq_vt}, we can see that if $t>0$, then $\bar{b}=v_{l+2}=v_{l}$. However, if $t=0$, we get no information from \Cref{eq_vt}. Furthermore, if $t$ is even, we have $v_{l+1}=v_{l-t+1}$, $v_{l+2}=v_{l-t+2}$; if $t$ is odd, we have $v_{l+1}=v_{l-t+2}$, $v_{l+2}=v_{l-t+1}$. Therefore, we must have $v_{l-t+1}\ne v_{l-t+2}$ since $v_{l+1}=b\ne\bar{b}=v_{l+2}$.
\par
Now we divide our discussions into two cases, according to \Cref{eq_case1} or \Cref{eq_case2}. As we will see later, they will lead to different values of $\bm{v}$.

\begin{itemize}
\item Firstly, we assume that \Cref{eq_case1} holds. Then $v_{s-1}=\tilde{x}_{s+1}=\alpha=\overline{\tilde{x}}_{s+2}=\bar{v}_s$, $v_{l-t}=\tilde{x}_{l+2-t}=\beta$, $v_{s+1}=\tilde{y}_{s+1}=\bar{\alpha}$, $v_{l+2-t}=\tilde{y}_{l+2-t}=\bar{\beta}=\overline{\tilde{y}}_{l+1-t}=\bar{v}_{l+1-t}$ and $\tilde{x}_{s+3}\cdots \tilde{x}_{l+1-t}=\widetilde{\bm{v}}=\tilde{y}_{s+2}\cdots\tilde{y}_{l-t}$. Thus we have
\begin{equation}\label{eq_xf1}
  \begin{array}{l}
    v_{s-1}=\bar{v}_s=\bar{v}_{s+1}, \\
    v_{l-t}=\bar{v}_{l-t+2}=v_{l-t+1}, \\
    v_{s+1}\cdots v_{l-t-1}=v_{s+2}\cdots v_{l-t}.
  \end{array}
\end{equation}
The three identities above imply that
\begin{equation}\label{eq_const}
  \bar{v}_{s-1}=v_s=\cdots =v_{l-t+1}=\bar{v}_{l-t+2}.
\end{equation}
Therefore, if $t=0$, then $\bar{b}=v_{l+2}=\bar{v}_l$.
Combining \Cref{eq_alterv,eq_vt,eq_const}, we get
\begin{equation}\label{eq_result1}
\bm{v}=v_1\cdots v_l=
\left\{
\begin{array}{ll}
  (a\bar{a})^{\frac{s}{2}}\bar{a}^{l-s-t}(\bar{a}a)^{\frac{t}{2}}, & \text{ }s,t \text{ both even}, \\
  (a\bar{a})^{\frac{s}{2}}\bar{a}^{l-s-t}(\bar{a}a)^{\frac{t-1}{2}}\bar{a}, & \text{ }s\text{ even},t\text{ odd},  \\
  (a\bar{a})^{\frac{s-1}{2}}a^{l+1-s-t}(a\bar{a})^{\frac{t}{2}}, & \text{ }s\text{ odd},t\text{ even}, \\
  (a\bar{a})^{\frac{s-1}{2}}a^{l+1-s-t}(a\bar{a})^{\frac{t-1}{2}}a, & \text{ }s,t \text{ both odd},
\end{array}
\right.
\end{equation}
for all $s,t\ge0$ and $l\ge s+t+1$. Recall that $v_l=\bar{b}$ if $t\ge 1$ and $v_l=b$ if $t=0$. So the first and last rows correspond to the case where $a=\bar{b}$, while the second and third rows correspond to the case where $a=b$.

\item Secondly, we assume that \Cref{eq_case2} holds. Comparing with the previous case, the only difference is that \Cref{eq_xf1} becomes
\begin{equation}\label{eq_xf2}
  \begin{array}{l}
    v_{s-1}=v_{s+2}=\bar{v}_{s+1}, \\
    \bar{v}_{l-t}=v_{l-t+2}=v_{l-t-1}, \\
    v_{s}\cdots v_{l-t-2}=v_{s+3}\cdots v_{l-t+1}.
  \end{array}
\end{equation}
Therefore, if $t=0$, we have $\bar{b}=v_{l+2}=\bar{v}_l$. Recall that $l\ge s+t+1$. From the first two rows in \Cref{eq_xf2}, we can deduce that $l\ge s+t+2$. Otherwise, the first two rows in \Cref{eq_xf2} will result in $v_{s-1}=\bar{v}_{s+1}=v_s$, which contradicts \Cref{eq_alterv}. Now the three identities in \Cref{eq_xf2} indicate that $v_{s+1}\cdots$ $v_{l-t+1}$ has period $3$, and that $v_{s+1}=\bar{v}_{s-1}=v_s=v_{s+3}$ and $v_{s+2}=v_{s-1}$.

If $l\equiv s+t\pmod{3}$, then $l-t+1\equiv s+1\pmod{3}$ and $l-t-1\equiv s+2\pmod{3}$. So from the periodicity of $v_{s+1}\cdots$ $v_{l-t+1}$, we have $v_{l-t+1}=v_{s+1}=v_s$ and $v_{l-t+2}=v_{l-t-1}=v_{s+2}=v_{s-1}\ne v_s=v_{l-t+1}$. Similarly,
if $l\equiv s+t+1\pmod{3}$, then $l-t+1\equiv (s+1)+1\pmod{3}$ and $l-t-1\equiv s+3\pmod{3}$. So $v_{l-t+1}=v_{s+2}=v_{s-1}$ and $v_{l-t+2}=v_{l-t-1}=v_{s+3}=v_{s}\ne v_{l-t+1}$. If $l\equiv s+t+2\pmod{3}$, then $l-t+1\equiv (s+1)+2\pmod{3}$. So $v_{l-t+1}=v_{s+3}=v_{s}$ and $v_{l-t+2}=\overline{v}_{l-t}=\overline{v}_{s+2}=\overline{v}_{s-1}=v_{s}$, where the last equality follows from \Cref{eq_alterv}. This contradicts the fact that $v_{l-t+2}\ne v_{l-t+1}$. So we can not have $l\equiv s+t+2\pmod{3}$.

By the above discussions, we have
\begin{equation}\label{eq_result2}
  v_{s+1}\cdots v_{l-t}=
  \begin{cases}
    (v_sv_{s-1}v_s)^{\frac{l-s-t}{3}}, & \mbox{if }l\equiv s+t\pmod{3},\\
    (v_sv_{s-1}v_s)^{\frac{l-s-t-1}{3}}v_s, & \mbox{if }l\equiv s+t+1\pmod{3}.
  \end{cases}
\end{equation}
Recall that $v_l=\bar{b}$ if $t\ge 1$ and $v_l=b$ if $t=0$.
Now, combining \Cref{eq_alterv,eq_vt,eq_result2}, for all $s,t,\ge 0$ and $l\ge s+t+2$, we get eight cases for the values of $\bm{v}$, which are listed in \Cref{tb_case2}. Note that in the third and seventh rows of \Cref{tb_case2}, we can not have $t=0$. Otherwise, we will get $v_l=v_{l-1}$, which contradicts the second row in \Cref{eq_xf2}.
\end{itemize}

From \Cref{eq_result1} and \Cref{tb_case2}, we can derive the results in \Cref{tb_thmC1}. Precisely  (in  \Cref{tb_case2}, we require $l\ge s+t+2$),
\begin{itemize}
  \item the first row in \Cref{tb_thmC1} follows from the second row in \Cref{eq_result1};
  \item the second row in \Cref{tb_thmC1} follows from the third row in \Cref{eq_result1};
  \item the third row in \Cref{tb_thmC1} follows from the second and third rows in \Cref{tb_case2};
  \item the fourth row in \Cref{tb_thmC1} follows from the fifth and eighth rows in \Cref{tb_case2};
  \item the fifth row in \Cref{tb_thmC1} follows from the first row in \Cref{eq_result1};
  \item the sixth row in \Cref{tb_thmC1} follows from the fourth row in \Cref{eq_result1};
  \item the seventh row in \Cref{tb_thmC1} follows from the first and fourth rows in \Cref{tb_case2};
  \item the eighth row in \Cref{tb_thmC1} follows from the sixth and seventh rows in \Cref{tb_case2}.
\end{itemize}
Therefore, we have prove the ``only if" part of the conclusion.

For the ``if" part, it is straightforward to verify that if we take $\bm{v}$ to be one of these values, $a\bar{a}\bm{v}$ and $\bm{v}b\bar{b}$ are Type-B confusable.
Comparing \Cref{tb_thmC1} with the results in (\Rnum{1}), we can see that $a\bar{a}\bm{v}$ and $\bm{v}b\bar{b}$ are not Type-A confusable. This completes the proof of (\Rnum{2}).
\begin{table}[t]
\centering
\caption{values of $\bm{v}$ derived from \Cref{eq_case2}}
\label{tb_case2}
\begin{tabular}{c|c|c|c}
  \hline
   No.&$\bm{v}$&conditions for $l,s$ and $t$& $a=b$\\
   \hline
   1&$(a\bar{a})^{\frac{s}{2}}(\bar{a}a\bar{a})^{\frac{l-s-t}{3}}(\bar{a}a)^{\frac{t}{2}}$&
   \tabincell{l}{$s$ is even, $t$ is even, \\ $l\equiv s+t\pmod{3}$} & No\\
   \hline
  2& $(a\bar{a})^{\frac{s}{2}}(\bar{a}a\bar{a})^{\frac{l-s-t}{3}}(\bar{a}a)^{\frac{t-1}{2}}\bar{a}$& \tabincell{l}{$s$ is even, $t$ is odd, \\ $l\equiv s+t\pmod{3}$} & Yes\\
   \hline
  3& $(a\bar{a})^{\frac{s}{2}}(\bar{a}a\bar{a})^{\frac{l-s-t-1}{3}}(\bar{a}a)^{\frac{t}{2}}\bar{a}$& \tabincell{l}{$s$ is even, $t\ge 2$ is even \\ $l\equiv s+t+1\pmod{3}$} &Yes\\
   \hline
   4&$(a\bar{a})^{\frac{s}{2}}(\bar{a}a\bar{a})^{\frac{l-s-t-1}{3}}(\bar{a}a)^{\frac{t+1}{2}}$& \tabincell{l}{$s$ is even, $t$ is odd \\ $l\equiv s+t+1\pmod{3}$} & No\\
   \hline
   5&$(a\bar{a})^{\frac{s-1}{2}}a(a\bar{a}a)^{\frac{l-s-t}{3}}(a\bar{a})^{\frac{t}{2}}$& \tabincell{l}{$s$ is odd, $t$ is even \\ $l\equiv s+t\pmod{3}$} & Yes\\
   \hline
   6&$(a\bar{a})^{\frac{s-1}{2}}a(a\bar{a}a)^{\frac{l-s-t}{3}}(a\bar{a})^{\frac{t-1}{2}}a$& \tabincell{l}{$s$ is odd, $t$ is odd \\ $l\equiv s+t\pmod{3}$}& No\\
   \hline
   7&$(a\bar{a})^{\frac{s-1}{2}}a(a\bar{a}a)^{\frac{l-s-t-1}{3}}(a\bar{a})^{\frac{t}{2}}a$& \tabincell{l}{$s$ is odd, $t\ge 2$ is even \\ $l\equiv s+t+1\pmod{3}$} &No\\
   \hline
   8&$(a\bar{a})^{\frac{s-1}{2}}a(a\bar{a}a)^{\frac{l-s-t-1}{3}}(a\bar{a})^{\frac{t+1}{2}}$& \tabincell{l}{$s$ is odd, $t$ is odd \\ $l\equiv s+t+1\pmod{3}$} & Yes\\
   \hline
\end{tabular}
\end{table}
\end{IEEEproof}

\section{Proofs of \Cref{lem_part1} and \Cref{lem_part2}}\label{appendix5}
\Cref{lem_part1} is the combination of \Cref{clm_part1}--\Cref{clm_part4}, and \Cref{lem_part2} is the combination of \Cref{clm_part1} and \Cref{clm_part5}--\Cref{clm_part7}.

In this section, we always denote $k=\left|\bm{w}\right|$, $s=\left|\bm{\alpha}\right|$ and $t=\left|\bm{\beta}\right|$. Then $k\ge 1$, $s\ge 0$ and $t\ge 0$. Let $\beta_{t+1}\triangleq\bar{b}$.
For simplicity, we define $\widetilde{\bm{x}}\triangleq a\bm{d}$ and $\widetilde{\bm{y}}\triangleq\bm{e}\bar{b}$. So $\left|\widetilde{\bm{x}}\right|=\left|\widetilde{\bm{y}}\right|=k+s+t$. Since $\left|S_{b}^{\bar{a}}\right|\ge 1$, we conclude that $\widetilde{\bm{x}}$ and $\widetilde{\bm{y}}$ are Type-A confusable or Type-B confusable. Therefore, we can choose $i$ to be the leftmost index such that $\widetilde{x}_i\ne\widetilde{y}_i$ and $j$ to be the rightmost index such that $\widetilde{x}_j\ne\widetilde{y}_j$. By the choice of $i$ and $j$, we have $\widetilde{x}_1\cdots \widetilde{x}_{i-1}$ $=$ $\widetilde{y}_1\cdots \widetilde{y}_{i-1}$ and $\widetilde{x}_{j+1}\cdots \widetilde{x}_{s+k+t}$ $=$ $\widetilde{y}_{j+1}\cdots \widetilde{y}_{s+k+t}$. Furthermore, according to \Cref{df_typA} and \Cref{df_typB}, we conclude that
\begin{itemize}
  \item if $\widetilde{\bm{x}}$ and $\widetilde{\bm{y}}$ are Type-A confusable, then $\widetilde{x}_i\cdots \widetilde{x}_{j}$ is an alternating sequence and $\widetilde{x}_i\cdots \widetilde{x}_j=\overline{\widetilde{y}_i\cdots \widetilde{y}_j}$.
  \item if $\widetilde{\bm{x}}$ and $\widetilde{\bm{y}}$ are Type-B confusable, then $j-i\ge 2$. Besides, one of the following two conditions must hold:
      \begin{equation}\label{eq_parttypB1}
        \widetilde{x}_i\ne\widetilde{x}_{i+1}=\widetilde{y}_i, \widetilde{x}_j=\widetilde{y}_{j-1}\ne\widetilde{y}_{j} \text{ and } \widetilde{x}_{i+2}\cdots \widetilde{x}_{j-1}=\widetilde{y}_{i+1}\cdots \widetilde{y}_{j-2},
      \end{equation}
      or
      \begin{equation}\label{eq_parttypB2}
        \widetilde{x}_i=\widetilde{y}_{i+1}\ne\widetilde{y}_{i}, \widetilde{x}_{j-1}=\widetilde{y}_{j}\ne\widetilde{x}_{j} \text{ and } \widetilde{x}_{i+1}\cdots \widetilde{x}_{j-2}=\widetilde{y}_{i+2}\cdots \widetilde{y}_{j-1}.
      \end{equation}
\end{itemize}

\begin{clm}\label{clm_part1}
  If $j\le s+k+1$, then $\bm{\beta}$ is a periodic sequence and its period is at most $2$. If $i\ge s$, then $\bm{\alpha}$ is a periodic sequence and its period is at most $2$.
\end{clm}
\begin{IEEEproof}
We only prove the conclusion for $\bm{\beta}$, since for $\bm{\alpha}$, the proof is similar.

When $0\le t\le 2$, it is clear that the period of $\bm{\beta}$ is $2$. So we assume that $t\ge 3$.
When $j\le s+k+1$, we have $\beta_1\cdots \beta_{t-1}$ $=$ $\widetilde{x}_{s+k+2}\cdots\widetilde{x}_{s+k+t}$ $=$ $\widetilde{y}_{s+k+2}\cdots\widetilde{y}_{s+k+t}$ $=$ $\beta_3\cdots \beta_{t}\bar{b}$, which implies that $\bm{\beta}$ is a periodic sequence and its period is at most $2$.
\end{IEEEproof}

\begin{clm}\label{clm_part2}
   Suppose that $i\ge s+k+2$. If $a\bm{d}$ and $\bm{e}\bar{b}$ are Type-A confusable, then $\bm{\beta}=\bm{p}_1\bm{p}_2$, where $\bm{p}_1,\bm{p}_2\in\Sigma_2^{*}$ are sequences of period at most $2$.
\end{clm}
\begin{IEEEproof}
Since $s+k+2\le i\le s+k+t$, we know that $t\ge 2$. When $t=2$, the conclusion is trivial. So we assume that $t\ge 3$.
When $i\ge s+k+2$, we have $\beta_1\cdots \beta_{i-s-k-2}$ $=$ $\widetilde{x}_{s+k+2}\cdots\widetilde{x}_{i-1}$ $=$ $\widetilde{y}_{s+k+2}\cdots\widetilde{y}_{i-1}$ $=$ $\beta_3\cdots \beta_{i-s-k}$, $\beta_{i-s-k-1}\cdots \beta_{j-s-k-1}$ $=$ $\widetilde{x}_{i}\cdots\widetilde{x}_{j}$ $=$ $\overline{\widetilde{y}_{i}\cdots\widetilde{y}_{j}}$ $=$ $\overline{\beta_{i-s-k+1}\cdots \beta_{j-s-k+1}}$ and $\beta_{j-s-k}\cdots \beta_{t-1}$ $=$ $\widetilde{x}_{j+1}\cdots\widetilde{x}_{s+k+t}$ $=$ $\widetilde{y}_{j+1}\cdots\widetilde{y}_{s+k+t}$ $=$ $\beta_{j-s-k+2}\cdots \beta_{t+1}$. The first and third equalities imply that $\beta_1\cdots \beta_{i-s-k}$ and $\beta_{j-s-k}\cdots \beta_{t+1}$ both have period at most $2$. Recall that $\widetilde{x}_i\cdots \widetilde{x}_{j}$ is an alternating sequence. So $j-i=0$ or $1$. Otherwise the second equality will lead to $\beta_{i-s-k+1}=\beta_{i-s-k-1}=\bar{\beta}_{i-s-k+1}$, which is impossible. Now we can conclude that $\beta$ is the concatenation of two sequences, and each of them has period at most $2$.
\end{IEEEproof}

\begin{clm}\label{clm_part3}
   Suppose that $i\le s+k+1$ and $j\ge s+k+2$. If $a\bm{d}$ and $\bm{e}\bar{b}$ are Type-A confusable, then $\bm{\beta}=\bm{p}_1\bm{p}_2$, where $\bm{p}_1,\bm{p}_2\in\Sigma_2^{*}$ are sequences of period at most $2$.
\end{clm}
\begin{IEEEproof}
Since $s+k+2\le j\le s+k+t$, we know that $t\ge 2$. When $t=2$, the conclusion is trivial. So we assume that $t\ge 3$.
When $i\le s+k+1$ and $j\ge s+k+2$, we have $\beta_1\cdots \beta_{j-s-k-1}$ $=$ $\widetilde{x}_{s+k+2}\cdots\widetilde{x}_{j}$ $=$ $\overline{\widetilde{y}_{s+k+2}\cdots\widetilde{y}_{j}}$ $=$ $\overline{\beta_3\cdots \beta_{j-s-k+1}}$ and $\beta_{j-s-k}\cdots \beta_{t-1}$ $=$ $\widetilde{x}_{j+1}\cdots\widetilde{x}_{s+k+t}$ $=$ $\widetilde{y}_{j+1}\cdots\widetilde{y}_{s+k+t}$ $=$ $\beta_{j-s-k+2}\cdots \beta_{t+1}$. The second equality implies that $\beta_{j-s-k}\cdots \beta_{t+1}$ has period at most $2$. Recall that $\widetilde{x}_i\cdots \widetilde{x}_{j}$ is an alternating sequence and $j\ge s+k+2$. So $j=s+k+2$ or $s+k+3$. Otherwise the first equality will lead to $\beta_{3}=\beta_{1}=\bar{\beta}_{3}$, which is impossible. Now we can take $\bm{p}_1=\beta_1\beta_2$ and $\bm{p}_2=\beta_3\cdots\beta_{t}$.
\end{IEEEproof}

Similar to \Cref{clm_part2} and \Cref{clm_part3}, we can prove the following claim.
\begin{clm}\label{clm_part4}
Suppose that $a\bm{d}$ and $\bm{e}\bar{b}$ are Type-A confusable, and that $s\ge3$.
If $j\le s-1$, or if $i\le s-1$ and $j\ge s$, then $\bm{\alpha}=\bm{p}_1\bm{p}_2$, where $\bm{p}_1,\bm{p}_2\in\Sigma_2^{*}$ are sequences of period at most $2$.
\end{clm}

\begin{clm}\label{clm_part5}
  Suppose that $i\le s+k+1$ and $j\ge s+k+2$, and that $a\bm{d}$ and $\bm{e}\bar{b}$ are Type-B confusable. Then
  \begin{itemize}
    \item $\bm{\beta}=\beta_1\bm{p}_1\bm{p}_2$, where $\bm{p}_1,\bm{p}_2\in\Sigma_2^{*}$ are sequences of period at most $2$; or
    \item $\bm{\beta}=\bm{p}_1\bm{p}_2$, where $\bm{p}_1,\bm{p}_2\in\Sigma_2^{*}$ are sequences of period at most $3$.
  \end{itemize}
\end{clm}
\begin{IEEEproof}
When $t\le 3$ the conclusion is trivial. So we assume that $t\ge 4$. If \Cref{eq_parttypB1} is true, then $\beta_2\cdots\beta_{j-s-k-2}$ $=$ $\widetilde{x}_{s+k+3}\cdots \widetilde{x}_{j-1}$ $=$ $\widetilde{y}_{s+k+2}\cdots \widetilde{y}_{j-2}$ $=$ $\beta_3\cdots\beta_{j-s-k-1}$ and $\beta_{j-s-k}\cdots\beta_{t-1}$ $=$ $\widetilde{x}_{j+1}\cdots \widetilde{x}_{s+k+t}$ $=$ $\widetilde{y}_{j+1}\cdots \widetilde{y}_{s+k+t}$ $=$ $\beta_{j-s-k+2}\cdots\beta_{t+1}$. Now we can conclude that $\beta_2\cdots\beta_{j-s-k-1}$ is a run and hence has period $1$, and that $\beta_{j-s-k}\cdots\beta_{t+1}$ is a sequence of period at most $2$.

If \Cref{eq_parttypB2} is true, then $\beta_1\cdots\beta_{j-s-k-3}$ $=$ $\widetilde{x}_{s+k+2}\cdots \widetilde{x}_{j-2}$ $=$ $\widetilde{y}_{s+k+3}\cdots \widetilde{y}_{j-1}$ $=$ $\beta_4\cdots\beta_{j-s-k}$ and
\newline
$\beta_{j-s-k}\cdots\beta_{t-1}$ $=$ $\widetilde{x}_{j+1}\cdots \widetilde{x}_{s+k+t}$ $=$ $\widetilde{y}_{j+1}\cdots \widetilde{y}_{s+k+t}$ $=$ $\beta_{j-s-k+2}\cdots\beta_{t+1}$. Now we can conclude that $\beta_1\cdots\beta_{j-s-k}$ is sequence of period at most $3$, and that $\beta_{j-s-k}\cdots\beta_{t+1}$ is a sequence of period at most $2$.
\end{IEEEproof}

\begin{clm}\label{clm_part6}
  Suppose that $i\ge s+k+2$, and that $a\bm{d}$ and $\bm{e}\bar{b}$ are Type-B confusable. Then $\bm{\beta}=\bm{p}_1\bm{p}_2\bm{p}_3$, where $\bm{p}_1,\bm{p}_2,\bm{p}_3\in\Sigma_2^{*}$ are sequences of period at most $3$.
\end{clm}
\begin{IEEEproof}
Similar to the proof of \Cref{clm_part2}, we can see that $\beta_1\cdots \beta_{i-s-k}$ and $\beta_{j-s-k}\cdots \beta_{t+1}$ both have period at most $2$.
If \Cref{eq_parttypB1} is true, then $\beta_{i-s-k+1}\cdots\beta_{j-s-k-2}$ $=$ $\widetilde{x}_{i+2}\cdots \widetilde{x}_{j-1}$ $=$ $\widetilde{y}_{i+1}\cdots \widetilde{y}_{j-2}$ $=$ $\beta_{i-s-k+2}\cdots\beta_{j-s-k-1}$. Now we can conclude that $\beta_{i-s-k+1}\cdots\beta_{j-s-k-1}$ is a run and hence has period $1$.

If \Cref{eq_parttypB2} is true, then $\beta_{i-s-k}\cdots\beta_{j-s-k-3}$ $=$ $\widetilde{x}_{i+1}\cdots \widetilde{x}_{j-2}$ $=$ $\widetilde{y}_{i+2}\cdots \widetilde{y}_{j-1}$ $=$ $\beta_{i-s-k+3}\cdots\beta_{j-s-k}$. Therefore, $\beta_{i-s-k}\cdots\beta_{j-s-k}$ is sequence of period at most $3$.
\end{IEEEproof}

Similar to \Cref{clm_part5} and \Cref{clm_part6}, we can prove the following claim.
\begin{clm}\label{clm_part7}
  Suppose that $a\bm{d}$ and $\bm{e}\bar{b}$ are Type-B confusable, and that $s\ge 4$.
  \begin{itemize}
      \item If $j\ge s$ and $i\le s-1$, then
           \begin{itemize}
              \item $\bm{\alpha}=\bm{p}_1\bm{p}_2\alpha_s$, where $\bm{p}_1,\bm{p}_2\in\Sigma_2^{*}$ are sequences of period at most $2$; or
              \item $\bm{\alpha}=\bm{p}_1\bm{p}_2$, where $\bm{p}_1,\bm{p}_2\in\Sigma_2^{*}$ are sequences of period at most $3$.
            \end{itemize}
      \item If $j\le s-1$, then $\bm{\alpha}=\bm{p}_1\bm{p}_2\bm{p}_3$, where $\bm{p}_1,\bm{p}_2,\bm{p}_3\in\Sigma_2^{*}$ are sequences of period at most $3$.
  \end{itemize}
\end{clm}

\section{The proof of \Cref{thm_tinsertion}}\label{appendix6}
Firstly, we need the following lemma, which is a generalization of the base case (i.e., the case where $\bm{u}=\bm{w}=\emptyset$) in the proof of \Cref{lem_basecase2}.
\begin{lem}\label{lem_basecase1}
  Let $n\ge 3$ and $\bm{x},\bm{y}\in\Sigma_2^n$ such that
  $$
  \bm{x}=a\bar{a}\bm{v}b,\text{  }\bm{y}=\bar{a}\bm{v}b\bar{b}
  $$
  for some $a,b\in\Sigma_2$ and $\bm{v}$ $\in\Sigma_2^{n-3}$. If $I_{1}(\bm{x})\cap I_{1}(\bm{y})=\{\bm{z}\}$, then
  $$
  \left|I_{t}(\bm{x})\cap I_{t}(\bm{y})\right|\le \left| I_{t-1}(\bm{z})\right|+N_2^{+}(n-1,t-1)=I_2(n+1,t-1)+N_2^{+}(n-1,t-1)
  $$
  for any $t\ge 2$.
\end{lem}
\begin{IEEEproof}
It is clear that $\bm{z}=a\bar{a}\bm{v}b\bar{b}$.
Let $S=I_{t}(\bm{x})\cap I_{t}(\bm{y})$. Then $S=S^{a}\cup S_{\bar{b}}\cup S_{b}^{\bar{a}}$ and
$$
\begin{array}{l}
  S^{a}=a\left(I_t(\bar{a}\bm{v}b)\cap I_{t-1}(\bar{a}\bm{v}b\bar{b})\right)=aI_{t-1}(\bar{a}\bm{v}b\bar{b})\subseteq I_{t-1}(\bm{z}), \\
  S_{\bar{b}}=\left(I_{t-1}(a\bar{a}\bm{v}b)\cap I_t(\bar{a}\bm{v}b)\right)\bar{b}=I_{t-1}(a\bar{a}\bm{v}b)\bar{b}\subseteq I_{t-1}(\bm{z}), \\
  S_{b}^{\bar{a}}=\bar{a}\left(I_{t-1}(a\bar{a}\bm{v})\cap I_{t-1}(\bm{v}b\bar{b})\right)b.
\end{array}
$$
If $a\bar{a}\bm{v}=\bm{v}b\bar{b}$, then $a=b$ and $\bm{v}=(a\bar{a})^m$, or $a=\bar{b}$ and $\bm{v}=(a\bar{a})^ma$ for some $m\ge 0$.  For both cases, $\bm{x}$ and $\bm{y}$ are Type-A confusable by \Cref{prop_summary} (2), which contradicts the fact that $\left|I_{1}(\bm{x})\cap I_{1}(\bm{y})\right|=1$. Therefore, $a\bar{a}\bm{v}\ne\bm{v}b\bar{b}$ and hence
 $\left|I_{t}(\bm{x})\cap I_{t}(\bm{y})\right|=|S|=|S^{a}\cup S_{\bar{b}}|+|S_{b}^{\bar{a}}|\le \left| I_{t-1}(\bm{z})\right|+N_2^{+}(n-1,t-1)$.
\end{IEEEproof}

\vspace{10pt}
\begin{pf}
Since $\left|I_{1}(\bm{x})\cap I_{1}(\bm{y})\right|=1$, by \Cref{lem_typB} we can assume that $\bm{x}=\bm{u}a\bar{a}\bm{v}b\bm{w}$ and $\bm{y}=\bm{u}\bar{a}\bm{v}b\bar{b}\bm{w}$ for some $a,b\in\Sigma_2$ and $\bm{u},\bm{v},\bm{w}\in\Sigma_2^{*}$. Let $n_0=|a\bar{a}\bm{v}b|$. Then $n\ge n_0$.
Let $S=I_{t}(\bm{x})\cap I_{t}(\bm{y})$.
  We prove this theorem by induction on $n$ and $t$.
\par
The base case is $n=n_0$ or $t=2$. When $n=n_0$, the conclusion follows from \Cref{lem_basecase1}. When $t=2$, the conclusion follows from \Cref{lem_basecase2}, since $I_2(n+1,1)+N_2^{+}(n-1,1)=n+5$.
Now suppose we have proved this theorem for all $n^{\prime}<n$ and $t^{\prime}<t$. Without loss of generality, we assume that $\bm{u}=c\bm{u}^{*}$ for some $c\in\Sigma_2$ and $\bm{u}^{*}\in\Sigma_2^{*}$. Clearly, $S=S^{\bar{c}}\cup S^{c}$, and
$$
\begin{array}{c}
  S^{\bar{c}}=\bar{c}\left(I_{t-1}(\bm{x})\cap I_{t-1}({\bm{y}})\right), \\
  S^{c}=c\left(I_t(\bm{u}^{*}a\bar{a}\bm{v}b\bm{w})\cap I_t(\bm{u}^{*}\bar{a}\bm{v}b\bar{b}\bm{w})\right).
\end{array}
$$
By the induction hypothesis, $|S^{\bar{c}}|\le I_2(n+1,t-2)+N_2^{+}(n-1,t-2)$ and $|S^{c}|\le I_2(n,t-1)+N_2^{+}(n-2,t-1)$. By Equations~(\ref{Nl=0}) and (\ref{Nl=1}), we can get the following two identities,
\begin{equation*}
\begin{array}{c}
  I_2(n+1,t-1)=I_2(n+1,t-2)+I_2(n,t-1),\\
  \text{ }N_2^{+}(n-1,t-1)=N_2^{+}(n-1,t-2)+N_2^{+}(n-2,t-1).
\end{array}
\end{equation*}
Then the proof is completed.
\end{pf}

\bibliographystyle{IEEEtran}
\bibliography{ref}
\end{document}